\documentclass[journal=mmbox,manuscript=communication,layout=twocolumn]{achemso}

\usepackage[version=3]{mhchem} 
\usepackage{achemso,amsmath,bm,multirow,relsize}

\author{Georgios G. Vogiatzis}
\author{Doros N. Theodorou}
\email{doros@central.ntua.gr}
\affiliation[NTUA]{School of Chemical Engineering, National Technical University of Athens, 
9 Heroon Polytechniou Street, Zografou Campus, GR-15780 Athens, Greece}

\title{Local Segmental Dynamics and Stresses in Polystyrene - \ce{C60} Mixtures}

\makeatletter
\let\thetitle\@title
\let\theauthor\@author
\makeatother

\makeatletter
\renewcommand\section{\@startsection{section}{1}{\z@}%
                                  {-3.5ex \@plus -1ex \@minus -.2ex}%
                                  {2.3ex \@plus.2ex}%
                                  {\normalfont\small\bfseries}}                                  
\makeatother

\makeatletter
\renewcommand\subsection{\@startsection{subsection}{1}{\z@}%
                                  {-3.5ex \@plus -1ex \@minus -.2ex}%
                                  {2.3ex \@plus.2ex}%
                                  {\normalfont\small\small\bfseries}}
\makeatother

\begin{document}
\begin{abstract} 
\begin{figure} 
\begin{center}
  \includegraphics[clip,width=1.0\linewidth] {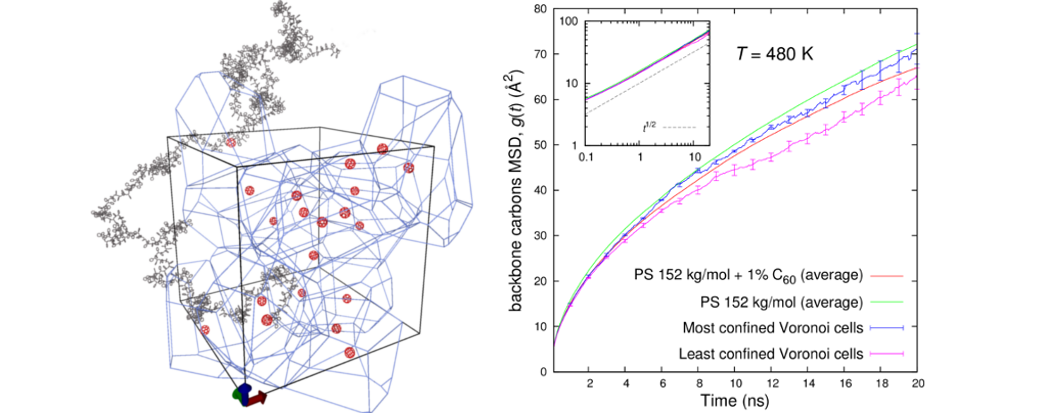}
\end{center}
\end{figure}

The polymer dynamics of homogeneous \ce{C60}-polystyrene mixtures in the molten state are studied via molecular 
simulations using two interconnected levels of representation for polystyrene nanocomposites: (a) A coarse-grained 
representation, in which each polystyrene repeat unit is mapped into a single ``superatom'' and each fullerene is 
viewed as a spherical shell. Equilibration of coarse-grained polymer-nanoparticle systems at all length scales is 
achieved via connectivity-altering Monte Carlo simulations. (b) An atomistic representation, where both nanoparticles 
and polymer chains are represented in terms of united-atom forcefields.
Initial configurations for atomistic Molecular Dynamics (MD) simulations are 
obtained by reverse mapping well-equilibrated coarse-grained configurations. 
By analyzing MD trajectories under constant energy, the segmental dynamics of polystyrene (for neat and filled systems) 
is characterized in terms of bond orientation time autocorrelation functions. Nanocomposite systems are found 
to exhibit slightly slower segmental dynamics than the unfilled ones, in good agreement with available experimental
data. Moreover, by employing Voronoi tessellation of the simulation box, 
the mean-squared displacement of backbone carbon atoms is quantified in the vicinity of each fullerene molecule. 
Fullerenes are found to suppress the average 
motion of polymeric chains, in agreement with neutron scattering data, while slightly increasing the dynamic and stress 
heterogeneity of the melt. Atomic-level and local (per Voronoi cell) stress distributions are reported for the pure 
and the filled systems. 
\end{abstract} 


\section{Introduction}
Nanomaterials fabricated by dispersing nanoparticles in polymer melts have the potential for performance that far
exceeds that of traditional composites. Nanoparticles have been shown to influence mechanical properties, as well 
as transport properties, such as viscosity. Until recently, the commonly held opinion was that 
particle addition to liquids, including polymeric liquids, produces an increase in viscosity, as predicted by Einstein
a century ago.\cite{AnnPhys_324_289,AnnPhys_339_591} However, it was recently found by Mackay and coworkers
\cite{NatMat_2_762,Macromolecules_38_8000,Macromolecules_40_9427} that the viscosity of polystyrene melts blended 
with crosslinked polystyrene particles (and later also with fullerenes and other particles) decreases and scales with 
the change in free volume (due to introduction of athermal excluded volume regions in the melt) and not with the 
decrease in entanglement. Later, \cite{Macromolecules_40_9427} fullerenes and magnetite particles were found to 
produce the same non-Einstein viscosity decrease effect.

Despite the macroscopic viscosity reduction, fullerene-polystyrene nanocomposites exhibit slower segmental dynamics, 
as Kropka et al.\cite{NanoLett_8_1061} have reported. The fully miscible \ce{C60} - polymer nanocomposites were made 
via a solution-dissolution / solvent-evaporation method. The molecular weight of the polystyrene chains was 
$M_{\rm w} = 156\;\text{kg/mol}$ with a narrow molecular weight distribution, $M_{\rm w}/M_{\rm n}=1.06$. 
The focus of their study was on nanocomposites containing a weight fraction $\phi_{\text{\ce{C60}}}^{\rm wt}=0.01$ 
because the most significant changes in $T_{\rm g}$ occur at this concentration. 
These materials exhibited an increase in their ``bulk'' $T_{\rm g}$ of about 1 K, as measured by differential 
scanning calorimetry and dynamic mechanical analysis. 
The mechanical measurements performed on these nanocomposites showed no evidence of excess structural or dynamic 
heterogeneity relative to the neat polymer and suggested that the effect of the particles may be described in terms of 
an increased segmental friction coefficient for the polymer. Quasi-elastic neutron scattering (QENS) 
measurements revealed that the influence of \ce{C60} on polymer melt dynamics is limited to the vicinity of the 
particle surfaces at the nanosecond time scale. 
The addition of \ce{C60}s was found to supress the polymer segmental dynamics of two more polymer hosts (PMMA and TMPC).
These authors suggested that the suppression of the local relaxation dynamics of the composite 
is consistent with an enhancement of cohesive interactions in the system, which may be the root of the increase 
in $T_{\rm g}$ for the nanocomposites. The system must acquire more thermal energy before polymer segments can 
overcome local energy barriers and thereby enable polymer center-of-mass motions. 
Specifically, local polymer chain backbone motions in the nanocomposites are 
suppressed relative to those in the neat polymer, an effect which likely plays a role in the observed 
increases in $T_{\rm g}$ of the materials. 
In the melt, the dynamics of the polymer segments in the vicinity of the particle surfaces is suppressed relative to 
the neat polymer, and this effect results in an excess elastic fraction of polymer segments at the nanosecond time 
scale. 

On the contrary, Sanz et al.\cite{JPhysCondsMatter_20_104209} reported increased segmental motion in 
polystyrene-fullerenes nanocomposites. These authors prepared bulk nanocomposites of (hydrogenous and ring-deuterated) 
polystyrene and poly(4-methyl styrene) using a rapid precipitation method where the \ce{C60} relative mass fraction
ranged from 0\% to 4\%. Elastic window scan measurements, using a high resolution ($0.9\;\mu{\rm eV}$) neutron 
backscattering spectrometer, were reported over a wide temperature range ($2-450\;{\rm K}$).
Based on the measured intensity, apparent Debye-Waller factors $\left\langle u^2\right\rangle$, characterizing the 
mean-squared amplitude of proton displacements, were determined as a function of temperature. 
Sanz et al. found that the addition of \ce{C60} to these polymers 
leads to a progressive increase in $\left\langle u^2\right\rangle$ relative to the pure polymer value over
the entire temperature range investigated, where the effect is larger for larger nanoparticle
concentration. This general trend seems to indicate that the \ce{C60} nanoparticles plasticize the fast dynamics of 
these polymer glasses. 

Later, Wong et al.,\cite{JMolLiq_153_79} investigated the same system (as Sanz et al.\cite{JPhysCondsMatter_20_104209}) 
by inelastic incoherent neutron scattering, small angle neutron scattering, calorimetric and dielectric spectroscopy 
methods. 
They found that the dispersion of fullerenes increased the glass transition temperature, slowing down the alpha 
relaxation dynamics associated with glass formation, while at the same time causing a softening of the material at 
high frequencies (as determined by the Debye-Waller factor). These effects are interpreted in terms of the 
particle modifying the polymer packing, causing an increase of the fragility of glass formation. 
The observed increase in $T_{\rm g}$ lies in apparent contradiction with the same groups's previous inelastic neutron 
scattering findings.\cite{JPhysCondsMatter_20_104209} 
From the point of view of these authors, this may indicate that the low-temperature slope of $\langle u^{2}\rangle$ 
increases with the addition of fullerene nanoparticles. The comparatively larger amplitude of proton delocalization 
in nanocomposites at the same temperatures is interpreted as a softening of the local potential of proton motion. 
In summary, inelastic neutron scattering indicates a simultaneous increase in amplitude of fast proton motion 
(increased mobility at fast time scales $\sim 10^{-15}\;{\rm s}$ and \AA-lengthscales), while restricted segmental 
motion associated with the glass transition is manifested by a $T_{\rm g}$ increase.

From the standpoint of molecular simulations, Vacatello\cite{Macromolecules_34_1946} performed Monte Carlo (MC) 
simulations of particles dispersed in a polymer matrix;
he found that polymer segments adhere to the particles and some chains are connected to different particles, thereby
forming ``bridges''. Each chain visits the interface layer of several particles and each particle can be in contact
with multiple chains. Even in the absence of strong interactions between particles and polymeric chains, Vacatello
observed that the particles behave as multifunctional physical crosslinks. These crosslinks do not immobilize the 
polymer chains, but can reduce their diffusion rates considerably. 

That view was complemented by molecular dynamics (MD) simulations by Desai et al.,\cite{JChemPhys_122_134910} who 
found that chain diffusivity is enhanced (relative to that in the pure polymer) when polymer-particle interactions 
are repulsive, and is reduced when polymer-particle interactions are strongly attractive. These authors were the 
first to report that chain diffusivity is spatially inhomogeneous, adopting its pure-polymer value when the chain 
center of mass is about one radius of gyration away from a particle's surface.
Smith et al.\cite{JChemPhys_117_9478} reported increase of the viscosity of a coarse-grained bead-necklace model 
polymer upon the addition of attractive and neutral nanoparticles, while viscosity reduction was observed upon the
addition of repulsive particles. 
Further MD simulations of polymer melts by Smith et al.\cite{PhysRevLett_90_226103} have suggested that both 
increased polymer segment packing densities and the energy topography of a surface can lead to stronger caging of 
polymer segments near an attractive surface. They suggested that the dramatic increase in structural relaxation 
time for polymer melts at the attractive structured surface is largely a result of dynamic heterogeneity induced by 
the surface and does not resemble dynamics of a bulk melt approaching $T_{\rm g}$.
The results of Kropka et al.\cite{NanoLett_8_1061} may indicate that \ce{C60}s induce similar effects in the glassy 
state of the polymers investigated.

More recently, Ndoro et al.\cite{Macromolecules_45_171} employed atomistic MD simulations in order to study the 
interfacial dynamics of 20-monomer atactic polystyrene chains surrounding a silica nanoparticle. 
The effect of the nanoparticle curvature and grafting density on the mean-squared displacement of free polystyrene 
chains and on the mean relaxation time of various intramolecular vectors was investigated as a function of 
separation from the surface. Confinement, reduced surface curvature, and densification resulted in a reduction of 
the mean-squared displacement and an increase in the mean relaxation time of the C-H bond vector and
chain end-to-end vector in the vicinity of the surface. 

Toepperwein et al.\cite{Macromolecules_44_1034} have addressed the influence of nanorods on the entanglement network 
of composites through MC and MD simulations. The presence of 
particles enriched the nanocomposite systems by nucleating additional topological constraints of polymer-particle
origin. Later, Li et al.\cite{PhysRevLett_109_118001} observed that highly entangled polymer chains were  
disentangled upon increasing the volume fraction of spherical non-attractive nanoparticles. 
The authors report a critical volume fraction controlling the crossover from polymer chain entanglements 
to nanoparticle-induced entanglements. While below this critical volume fraction, the polymer chain relaxation 
accelerates upon filling, above this volume fraction, the situation reverses. 
The same authors\cite{Macromolecules_45_2099} have also studied the structural, dynamical and viscous behaviors of 
polyethylene matrices under the influence of differently shaped nanoparticles, including \ce{C60} buckyballs, 
at fixed volume fraction (4 vol \%). The nanoparticles were found to be able to nucleate polymer entanglements around 
their surfaces and to increase the underlying entanglement density of the matrix in the proximity of the particles.
However, the overall primitive path networks of PE matrices were found very similar to those of pure polyethylene, 
since polymer entanglements still dominated the rheological behavior.

In this work we are trying to understand the microscopic mechanisms involved in the peculiar behavior of PS-\ce{C60}
nanocomposites, through a hierarchical simulation approach. To this end, we focus our study on polystyrene melts
having specifications identical to those studied by Kropka et al.\cite{NanoLett_8_1061}
The computational prediction of physical properties is particularly challenging for polymeric materials, because of 
the extremely broad spectra of length and time scales governing structure and molecular motion in these materials. 
This challenge can only be met through the development of hierarchical analysis and simulation strategies encompassing 
many interconnected levels, each level addressing phenomena over a specific window of time and 
length scales.\cite{ChemEngSci_62_5697}
In order to shed some light into the segmental dynamics of PS-\ce{C60} systems, molecular simulations have been conducted 
using two interconnected levels of representation for polystyrene nanocomposites: (a) A coarse-grained 
representation,\cite{JPhysChem_B_109_18609} in which each polystyrene repeat unit is mapped into a single ``superatom'' 
and each fullerene is viewed as a spherical shell. 
Equilibration of coarse-grained polymer-nanoparticle systems at all length scales is 
achieved via connectivity-altering Monte Carlo (MC) simulations.\cite{Macromolecules_40_3876} 
(b) An atomistic representation, where both 
nanoparticles and polymer chains are represented in terms of united-atom forcefields. Initial configurations 
for atomistic Molecular Dynamics (MD) simulations are obtained by reverse mapping well-equilibrated coarse-grained 
configurations. The reverse mapping procedure retains the tacticity which is implicit in the coarse-grained 
representation, while regrowing atomistic sites by a quasi-Metropolis procedure that avoids unphysical conformations. 
By analyzing MD trajectories under constant energy, the segmental dynamics of polystyrene (for neat and 
filled systems) can then be characterized in terms of bond orientation time autocorrelation functions and 
atomic local mean-squared displacement.

\section{Coarse Grained Monte Carlo (CG-MC)}

\subsection{Systems studied} 
In this work, monodisperse melts of atactic PS chains with 50 \% meso diads obeying Bernoullian statistics and with 
chain lengths of 80 ($8.3\:{\rm kg/mol}$), 185 ($19.3\:{\rm kg/mol}$), 323 ($34\:{\rm kg/mol}$), 
922 ($96\:{\rm kg/mol}$), 1460 ($152\:{\rm kg/mol}$) and 4032 ($420\:{\rm kg/mol}$) coarse-grained sites (diads) 
were generated and simulated.
Initial configurations were generated and equilibrated at the temperature of 500 K, at which the CG forcefield has been 
developed. Then the reverse-mapped well-equilibrated configurations were subjected to MD runs which cooled them down
to the glass transition temperature.
Bearing in mind the limitations of the atomistic MD simulations, the main part of this work is based on systems 
composed of $n = 10$ chains of molecular weight $152\:{\rm kg/mol}$ with 20 fullerenes randomly dispersed, 
leading to a weight fraction $\phi_\text{\ce{C60}}^{\rm wt} = 1\text{\%}$. 
System specifications are close to 
experimentally studied systems by quasi-elastic neutron scattering\cite{NanoLett_8_1061} and adequate for avoiding 
finite-size effects.\cite{JChemPhys_99_6983}
For comparison, neat polymeric systems of the same characteristics have also been simulated along with the composite ones. 
In order to improve statistics, three independent configurations of both the neat and the filled system were generated, 
equilibrated and reverse-mapped.
Moreover, a single 80-mer chain system was atomistically built and simulated with MD, without involving MC equilibration.

The major contribution to the computational cost of our study stems from the MD simulations. Despite the fact that our 
MC and reverse mapping codes are running serially, the time needed to fully equilibrate our systems and generate the 
corresponding atomistic configurations is reasonable (10-15 days of wall-clock time on a single 2.8 GHz CPU core).
On the contrary, the MD simulations of the systems described above required 40 days of real wall-clock time for 
170 ns integration time, when run on a hybrid machine consisting of eight 2.8 GHz CPU cores and one programmable 
graphics processing unit. Since the polymer relaxation times (e.g., the disentanglement or reptation time, $\tau_{\rm d}$) 
far exceed our current computational resources, the use of MC to equilibrate our samples is of vital importance.

\subsection{Coarse Grained model}

The coarse graining scheme adopted in this work is very efficient for the representation of vinyl polymers, 
since one is able to keep information on stereochemical sequences along the polymer chain. 
Given a direction along the main chain, it is possible to assign an absolute configuration to each asymmetric carbon. 
The chain can be represented as a sequence of diads, 
each diad containing two asymmetric carbons. Depending on the absolute configuration of the asymmetric carbons, i.e.,
\textit{RR} (or \textit{SS}) and \textit{RS} (or \textit{SR}), the diads can be of type \textit{m} (\textit{meso}) or
\textit{r} (\textit{racemo}), respectively (Figures 1 and 2 of ref \citenum{JPhysChem_B_109_18609}). The chain ends 
can be either \textit{em} (\textit{end-meso}) or \textit{er} (\textit{end-racemo}) as far as the bonded potentials are 
concerned, but their nonbonded interaction is common and slightly different from \textit{m} and \textit{r} non-bonded
interactions. A detailed description of the model can be found in ref \citenum{JPhysChem_B_109_18609} and the 
parameter values used are taken from the Supporting Information of ref \citenum{Macromolecules_40_3876}. 
They were derived from a detailed atomistic potential using the Iterative Boltzmann Inversion (IBI) method.
The coarse-grained effective potential was refined in order to better reproduce the target distributions, extracted from 
all-atom simulations of a 9mer fluid at 500 K and 1 bar. All MC simulations of the present work have been conducted at 
the same temperature.

At the coarse grained level of description, fullerenes are considered as spherical shells of infinitesimal
thickness. It is assumed that carbon atoms are uniformly smeared over the surface of the shell. The potential between
a spherical shell of interaction sites and a single coarse grained PS site (treated as a point), 
on the grounds of Hamaker theory,\cite{Physica_4_1058} is: 
\begin{align}
\mathcal{V}_{\text{\ce{C60}}}^{\text{CG}} (r) = & 8 \pi \epsilon_{\rm m}
\rho_{\text{\ce{C60}}} \delta r_{\text{\ce{C60}}} \nonumber \\
& \left[  
\frac{\sigma_{\rm m}^{12}}{10 r} \left(\frac{R_{\text{\ce{C60}}}}{\left(r + R_{\text{\ce{C60}}} \right)^{10}}
- \frac{R_{\text{\ce{C60}}}}{\left(r - R_{\text{\ce{C60}}} \right)^{10}} \right)  \right .
\nonumber \\
& - \left.  \frac{\sigma_{\rm m}^6}{4r}
\left(\frac{R_{\text{\ce{C60}}}}{\left(r + R_{\text{\ce{C60}}} \right)^4}
- \frac{R_{\text{\ce{C60}}}}{\left(r - R_{\text{\ce{C60}}} \right)^4} \right) \right]
\end{align}
where $r$ is the center-to-center separation distance. The external radius of the shell representing the fullerene, 
$R_\text{\ce{C60}}$, is set to $0.35\:{\rm nm}$. The density of interaction sites of fullerene is 
$\rho_\text{\ce{C60}}$, the thickness of the shell $\delta r_{\text{\ce{C60}}}$, and the product 
$\rho_{\text{\ce{C60}}} \delta r_{\text{\ce{C60}}}$ represents the surface density of interaction sites of the 
fullerene. The mean Lennard-Jones interaction parameters, $\epsilon_{\rm m}$ and $\sigma_{\rm m}$, take into account
the interaction of fullerene carbons with all kinds of interaction sites a coarse-grained bead consists of.
In the case of PS, for every bead we sum up five different kinds of interactions, namely the interaction of \ce{CH3},
aliphatic \ce{CH2}, aromatic CH and C groups with the fullerene. Each interaction is weighted with the number of 
interaction sites of each kind present in the coarse-grained (CG) bead. 
The detailed united-atom forcefield, which is essential for the calculation of $\epsilon_{\rm m}$ and
$\sigma_{\rm m}$, is described in the  ``Target atomistic representation'' subsection, which follows.
In \ref{tab:cg_mc_parameters} are listed the nonbonded interaction parameter values of the united-atom forcefield, 
as well as coarse-grained model parameters derived from them and used in this work.

Following the work of Girifalco, \cite{JPhysChem_96_858} the interaction between fullerenes at the coarse-grained level 
is modeled as an integrated Lennard-Jones potential over two spherical shells:
\begin{align}
\mathcal{V}_{\text{\ce{C60}-\ce{C60}}}(r) = -& \mathcal{A} \left(\frac{1}{s(s-1)^3} 
+ \frac{1}{s(s+1)^3} - \frac{2}{s^4}\right) \nonumber \\
+& \mathcal{B} \left(\frac{1}{s(s-1)^9} + \frac{1}{s(s+1)^9} - \frac{2}{s^{10}}\right)
\end{align}
where $s = r / \left(2 R_{\text{\ce{C60}}}\right)$ and the values of the remaining parameters are presented in  
\ref{tab:cg_mc_parameters}. 
This potential has been used throughout the literature to study the molecular properties of \ce{C60}, which were
found to be consistent with available experimental data, making it a reasonable choice.
A priori, there should be an entropic contribution included in the coarse-grained polymer-particle 
potentials,\cite{PhysChemChemPhys_13_10491} which is not taken into account by using Hamaker-type potentials.
However, our coarse-graining is relatively modest, so we believe that this contribution is small.
As our PS coarse-graining serves mainly for creating initial configurations, any fault in the local 
structure caused by the coarse-grained potential, will be fixed by the subsequent atomistic MD simulation.

\begin{table*}
	\centering
	\caption{United-atom potential interaction parameters and coarse-grained model parameters derived from them.}
	\label{tab:cg_mc_parameters}
	\begin{tabular}{lcc}
		\hline
		Parameter & Value & Reference \\ \hline
        \textit{Polystyrene} & & \\
		$\epsilon_{\text{\ce{CH3}}},\epsilon_{\text{al\ce{CH2}}},\epsilon_{\text{arCH}}
        ,\epsilon_{\text{arC}}$ & $0.12 \;\text{kcal/mol}$ & 
        \multirow{4}{*}{Mondello et al.\cite{Macromolecules_27_3566}} \\
        $\epsilon_{\text{al\ce{CH}}}$ & $0.09 \;\text{kcal/mol}$  &   \\
        $\sigma_\text{alCH}, \sigma_\text{arC}, \sigma_\text{arCH}$ & $3.69\;\text{\AA}$ & \\
        $\sigma_\text{\ce{CH3}}, \sigma_\text{al\ce{CH2}}$ & $3.85\;\text{\AA}$ & \\ \cline{1-1}
        \textit{Fullerene} & & \\
		$\epsilon_{\text{\ce{C60}}}$ & $0.066 \;\text{kcal/mol}$ & \multirow{4}{*}{Girifalco\cite{JPhysChem_96_858}} \\ 
		$\sigma_{\text{\ce{C60}}}$ & $3.47 \text{\AA}$ & \\
        $\mathcal{A}$ & $1.079 \;\text{kcal/mol}$ & \\
		$\mathcal{B}$ & $1.957 \times 10^{-3} \;\text{kcal/mol}$ & \\ \cline{1-1}
        \textit{Mixed} & & \\
        $\epsilon_{\rm m}$ & $\sqrt{\epsilon_\text{PS} \varepsilon_{\text{\ce{C60}}}}$ & 
        where ``PS'' subscript represents \\
		$\sigma_{\rm m}$ & $1/2\left(\sigma_\text{PS} + \sigma_\text{\ce{C60}}\right)$ & all united-atom species \\ 
      \hline
	\end{tabular}
\end{table*}

\subsection{Initial CG Structure Generation} 

To start the coarse-grained Monte Carlo simulation, an initial configuration is generated by placing the fullerenes
at randomly selected positions, so that they do not overlap, and then building stepwise the polymeric chains around 
them, following the work of Theodorou and Suter.\cite{Macromolecules_18_1467}
At each step a site type was chosen according to the overall probability of \textit{meso} and \textit{racemo} diads. 
In our work the tacticity of atactic PS is represented by a Bernoullian diad distribution with a probability of 
\textit{meso} diads of $(m)=0.5$ (in equilibrium atactic polystyrene the configuration statistics is almost 
Bernoullian with the fraction of \textit{meso} dyads around 0.46 
\cite{MakromolChemRapidCommun_3_181,MakromolChemRapidCommun_3_661,JPolymSciPolymPhysEd_21_1667}). 
The ``Euler angles'' defining the direction of the first and the second coarse-grained bonds, are arbitrarily set 
at the beginning. Since torsion angle potentials are not used in the CG representation, for each subsequent superatom
a bond angle is chosen according to a probability resulting from the effective bending potential for the specific angle 
type and the segment is placed, accordingly, on a circle forming the base of a cone with apex at the previously 
placed superatom and side length equal to the average CG bond length of $l_{\rm CG}=2.46\:\text{\AA}$. In every step, 
non-bonded interactions with already created superatoms are taken into account, refining the 
probabilities of accepting a trial position for the bead to be grown.
If, after a certain number of iterations, all attempts to grow the superatom fail, a local derivative-free minimization of 
the potential energy is undertaken in order to ensure that the bead is placed at the most energetically favorable position.
Since the degrees of freedom of the optimization are only three, i.e. the cartesian coordinates of the new 
superatom to be built, a numerically stable Nelder-Mead algorithm is a reasonable choice.\cite{ComputerJournal_7_308}

Each initial guess structure is then ``relaxed'' to a state of minimal potential energy. The total potential energy is 
the sum of all bonded potentials of the polymeric matrix, non-bonded interactions between polymer superatoms, between 
polymer superatoms and fullerenes, and between fullerenes and fullerenes. Minimization is carried out using the Large-scale 
Atomic/Molecular Massively Parallel Simulator (LAMMPS)\cite{JCompPhys_117_1} with the coarse-grained potentials 
incorporated in its source code. Since IBI PS coarse-grained non-bonded potentials are tabulated, a Polak-Ribi\`ere 
\cite{polak1971computational,RevFrancInformatRechOperationelle_16_35} variant of the conjugate gradient method is used.

\subsection{Monte Carlo equilibration}

Enabling the equilibration of high molar mass polymer nanocomposites at the coarse-grained level was one of the main 
objectives of this work. 
For this purpose, connectivity altering moves, \cite{PhysRevLett_88_105503} such as double bridging (DB) 
\cite{JChemPhys_117_5465} were employed in Monte Carlo 
simulations of linear chains of four types of sites, \textit{m}, \textit{r}, \textit{em} and \textit{er}. In the DB
move two trimers are excised from two chains of equal length and two new trimer bridges are constructed, leading to 
two new chains of the same length but of drastically different conformations. 
The internal shape of the chains was rearranged by using the symmetric variant of the concerted rotation move, 
\cite{MolPhys_78_961,Macromolecules_28_7224} which modifies the local conformation of an internal chain section
(of five superatoms) while leaving the preceding and following parts of the chain unaffected.  
In addition, the internal conformations of chains are sampled using flips of internal beads, end segment rotations and
reptations.
The mixture of MC moves included a newly developed and specially adapted configurational bias 
move,\cite{JChemPhys_96_2395} capable of regrowing an arbitrary number of coarse-grained beads, starting from a chain 
end, in a confined environment formed by the nanoparticles. 
For every kind of move undertaken, special care was taken to discard moves leading to overlaps of coarse-grained beads 
and dispersed nanoparticles.

\subsection{Conformational properties of CG configurations}
\begin{figure}
   \includegraphics[width=0.45\textwidth]{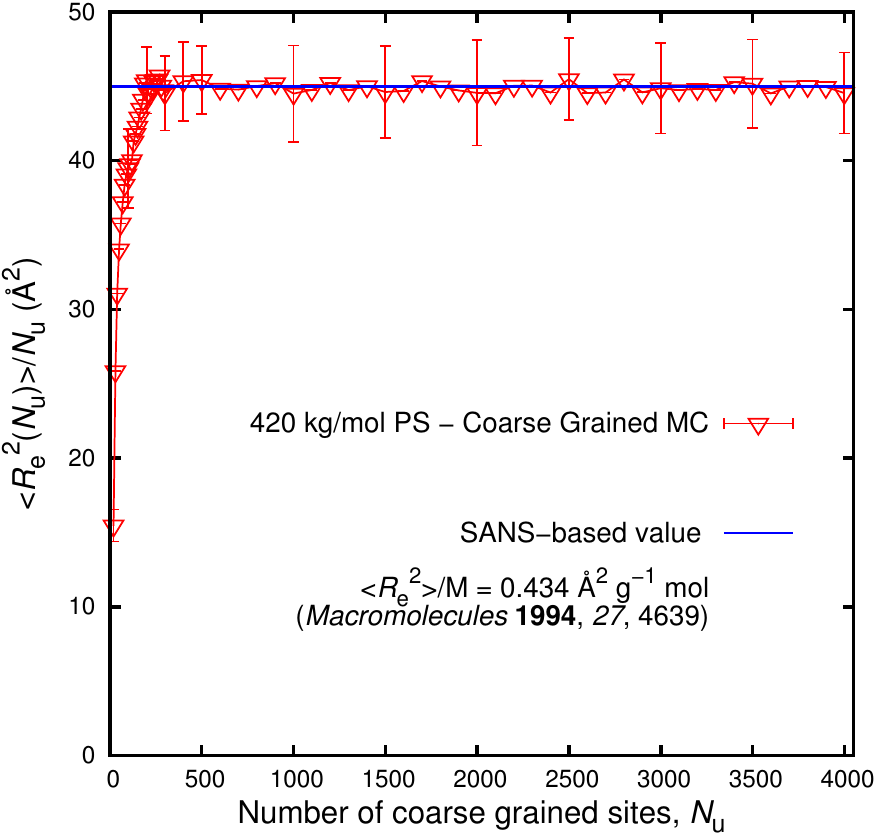}
   \caption{Average squared end-to-end distance, $\left\langle R_{\rm e}^2 (N_{\rm u}) \right\rangle$, of subchains 
            of length $N_{\rm u}$ diads divided by $N_{\rm u}$ versus $N_{\rm u}$ for coarse-grained 4000-mer chains
            at $500\:\text{K}$. Some indicative error bars are shown. The dotted line is the SANS-based value for 
            PS at $413\:\text{K}$.\cite{Macromolecules_27_4639}}
   \label{fig:ete_vs_nu}
\end{figure}

As a measure of the obtained chain conformations one may use the quantity 
$\left\langle R_{\rm e}^2(N_{\rm u})\right\rangle/N_{\rm u}$, where $N_{\rm u}$ is the number of diads (repeat 
units) in a subchain and $R_{\rm e}^2(N_{\rm u})$ is the squared end-to-end distance of the subchain. It has been 
shown that,\cite{JChemPhys_119_12718} for well equilibrated chains, 
$\left\langle R_{\rm e}^2(N_{\rm u})\right\rangle/N_{\rm u}$ increases monotonically with $N_{\rm u}$ until it reaches
a plateau. In  \ref{fig:ete_vs_nu}, $\left\langle R_{\rm e}^2(N_{\rm u})\right\rangle/N_{\rm u}$ is depicted as a 
function of $N_{\rm u}$ for a system containing 8 chains of 4000 diads each, at $T=500\:\text{K}$. For $N_{\rm u}>200$ 
an asymptotic value is reached of approximately $45\:\text{\AA}$ which corresponds to 
$\left\langle R_{\rm e}^2\right\rangle/M$ equal to $0.43\:\text{\AA}^{2}{\rm g}^{-1}{\rm mol}$, with $M$ being the 
chain molar mass. This is in excellent agreement with the Small Angle Neutron Scattering (SANS)-based value (shown 
with the dotted line) of $\left\langle R_{\rm e}^{2}\right\rangle = 0.434 \:\text{\AA}^{2}{\rm g}^{-1}{\rm mol}$
given for PS at $T=413\:{\rm K}$.\cite{Macromolecules_27_4639}

\begin{figure}
   \includegraphics[width=0.45\textwidth]{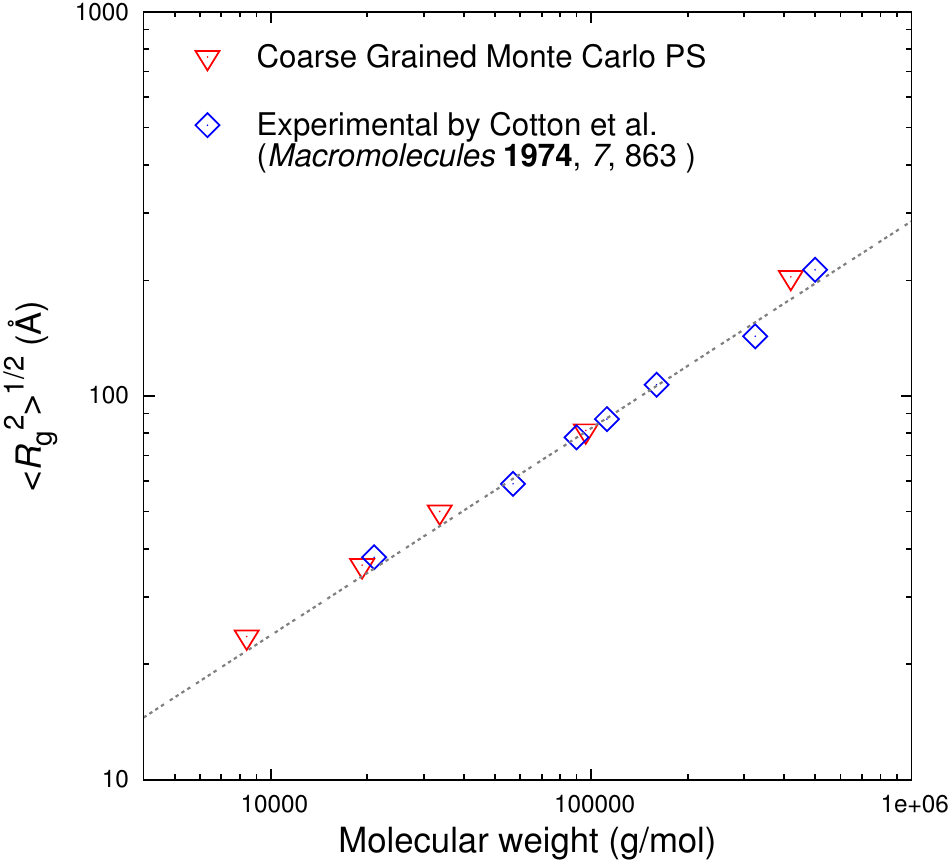}
   \caption{Root mean squared radius of gyration of the coarse grained chains as a function of molar mass, $M$,
            in the melt at $500\:\text{K}$ (triangular symbols).
            Neutron scattering measurements for high molar mass PS are also included (rhomboid 
            symbols).\cite{Macromolecules_7_863} Grey dotted line is linear least-squares fit.}
   \label{fig:rg_vs_mw}
\end{figure}

The calculated root-mean-square radius of gyration $\left\langle R_{\rm g}^2\right\rangle^{1/2}$ as a function of $M$
is shown in  \ref{fig:rg_vs_mw}. Neutron scattering results \cite{Macromolecules_7_863} for monodisperse
PS of $M$ ranging from $21\:\text{kg/mol}$ to $1100\:\text{kg/mol}$ in the bulk at $393\:\text{K}$ have 
also been drawn (continuous thin line) for comparison over the $M$ scale of interest. Very good agreement
is observed for all the molar masses examined.
This confirms that chains in our CG melts are well equilibrated and adopt close to unperturbed configurations.
At the CG level of representation, nanoparticles do not seem to affect the dimensions of the chains, yielding identical 
results with the neat polystyrene systems. However, this cannot be considered as a generally valid statement for polystyrene
nanocomposite melts (e.g. polystyrene chains have been found to swell upon the addition of crosslinked polystyrene 
nanoparticles \cite{EurPolymJ_47_699}).

\section{Reverse Mapping}

\subsection{Target atomistic representation}  
The equilibrated coarse-grained configurations are reverse mapped to detailed configurations which can be described by
a united-atom model without partial charges, based on the works of Mondello et al. \cite{Macromolecules_27_3566} 
and Lyulin and Michels.\cite{Macromolecules_35_6332}
This united-atom model will also be referred to as ``atomistic'' in the following.
It takes into account the following contributions to the system potential energy: (i) Lennard-Jones non-bonded 
interaction potential between all united atoms that are three or more bonds apart or belong to different images of 
the parent chain; (ii) bond stretching potential for every covalent bond; (iii) bending potential for all bond angles, 
including those in the phenyl rings; (iv) torsional potential for all rotatable backbone bonds; (v) torsional potential 
for the torsions of phenyl rings around their stems; (vi) out-of-plane bending potential to preserve the coplanarity 
of the phenyl and the phenyl stem; (vii) torsional potential about all bonds connecting aromatic carbons in the phenyl 
ring to preserve the planarity of the ring and (viii) improper torsional potential to preserve the chirality of all 
carbons bearing a phenyl substituent.\cite{JChemPhys_112_9632} 
Lennard-Jones parameters employed by the model are listed in \ref{tab:cg_mc_parameters}.
All Lennard-Jones potentials are cut at an inner cutoff distance of $2.35\sigma$, beyond which force smoothing to zero 
using a cubic spline is applied up to a distance of $2.5\sigma$.
No tail corrections are used for the non-bonded interaction potential.
Our experience has been that this united-atom model does a reasonable job predicting structure, volumetric properties,
elastic constants and stress-strain behavior in the melt and glassy state. In particular, the system adopts 
density values close to those measured experimentally upon quenching into the glassy state.\cite{MolPhys_111_3430}

Fullerenes are described as fully flexible carbon cages. 
As far as intramolecular (bond stretching, bending and torsional) contributions are concerned, the DREIDING forcefield 
has been used \cite{JPhysChem_94_8897}.
Intramolecular non-bonded interactions are not taken into account, while the intermolecular non-bonded interactions 
are described by a Lennard-Jones potential using the values of $\varepsilon_{\text{\ce{C60}}}$ and 
$\sigma_{\text{\ce{C60}}}$ reported in \ref{tab:cg_mc_parameters}, following the early but well validated work of 
Girifalco.\cite{JPhysChem_96_858}
A comprehensive review of \ce{C60} forcefields can be found in the work of Monticelli \cite{JChemTheoryComput_8_1370}.
In order to ensure the reliability of our atomistic MD simulations, all sets of parameters reported by Monticelli have 
been tested, rendering indistinguishable results, as far as the trajectories and the thermodynamic properties  
of the systems were concerned. 

\subsection{Procedure}
The reconstruction of the atomistic detail, given a well-equilibrated coarse-grained configuration, is accomplished in 
four stages. During the first stage, atomistic fullerenes are placed at the positions of the coarse-grained beads 
used during the CG-MC equilibration, while paying attention to select the orientation that minimizes the interaction 
energy with their environment. The second stage consists of an iterative quasi-Metropolis introduction of the atomistic 
sites of the polymer, obeying the atomistic potential described above. During the third stage, local MC moves try to optimize the 
generated configuration. At the final stage, energy minimization is undertaken before initiating the MD integration.
Throughout the atomistic reconstruction procedure, \ce{CH2} united atoms containing the achiral carbons of the chains 
are kept fixed at the positions of the superatoms of the coarse-grained configuration. 

In order to restore the atomistic detail of the coarse-grained PS, a quasi-Metropolis procedure is followed.
\cite{Macromolecules_18_1467} 
During the reconstruction of the atomistic sites, the positions of the united atoms added (aliphatic \ce{CH} groups
containing the chiral carbons of the chains, aromatic C and \ce{CH} groups constituting the phenyl substituents) are 
selected from a set of properly created candidates, using as a criterion the increase in the total energy of the system. 
The first chiral carbon of each chain was assigned randomly an absolute configuration, since this is not determined 
by the coarse-grained model, and the rest of the chiral carbons were placed according to the chirality of the diads.
Starting from the \ce{CH2} sites, whose position vectors are the degrees of freedom of the coarse-grained representation, backbone 
\ce{CH} united atoms are selected from a set of candidate positions lying in the circle formed by the intersection of 
two spheres, one centered at the previous \ce{CH2} atom and the other centered at the next one. The radii of the 
spheres correspond to the equilibrium length of the backbone C-C bond. 
Based on the positions of the aliphatic \ce{CH2} and \ce{CH} united atoms, the carbon defining the stem of the phenyl
ring is placed so that the aliphatic \ce{CH} - aromatic C bond generates close to equilibrium bending angles.
Finally, the rings are introduced as planar objects, the plane of each ring containing the axis of its stem, using as 
the only degree of freedom of the candidate positions the torsion angle of the ring around its stem.
During the whole regrowth, insertions of carbon atoms leading to a 
\textit{\=g} (\textit{gauche-bar})\cite{JAmChemSoc_96_5015} 
conformation are strictly prohibited by assigning to them zero probability. 

\begin{figure}
	\includegraphics[width=0.45\textwidth]{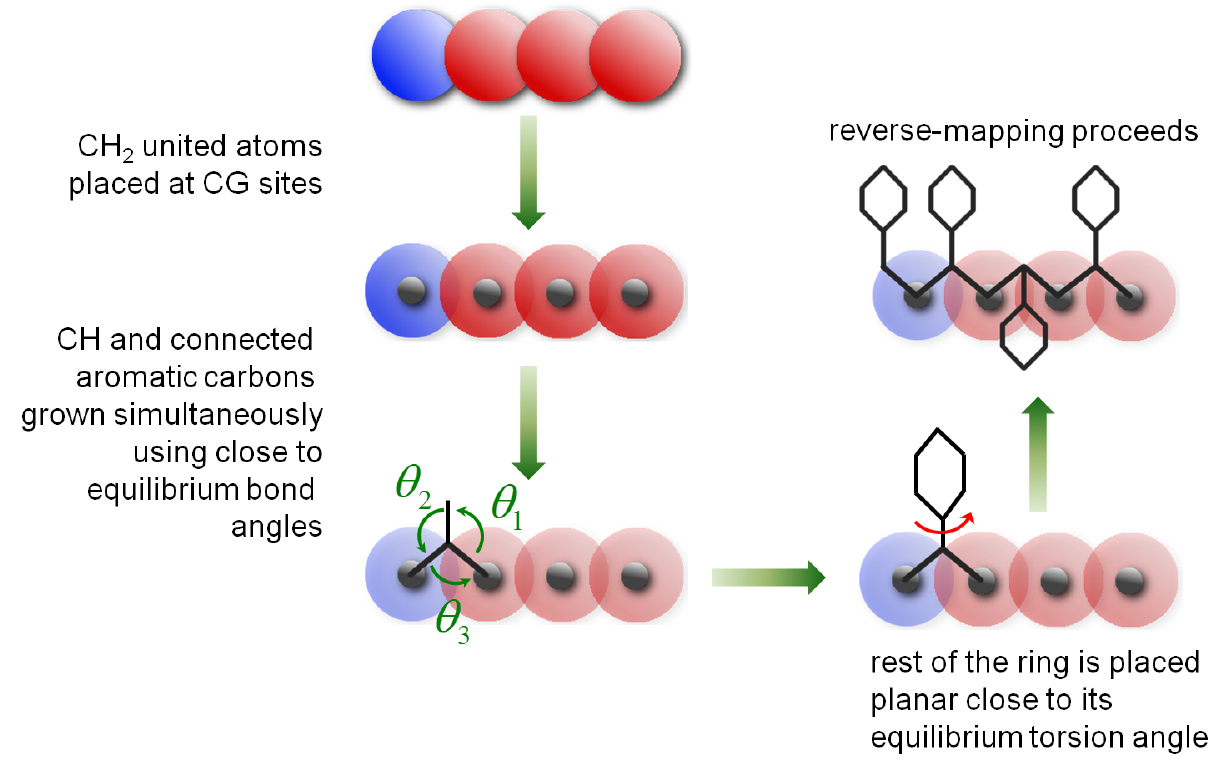}
	\caption{Schematic of the initial reconstruction of the atomistic detail.}
	\label{fig:rm_scheme}
\end{figure}

Following Spyriouni et al.,\cite{Macromolecules_40_3876} the configuration is optimized via local Monte Carlo moves. 
These moves include flip of a segment, rotation of the phenyl ring around its stem and configurationally biased 
regrowth of a whole monomer, preserving the chirality of the CG site. 
A flip move displaces an inner skeletal segment of the chain along the locus (circle) defined by the lengths of the two 
bonds adjacent to the segment. The moves employed flipped a chiral carbon (one carrying a phenyl) to a new position on 
the circumference of a circle drawn perpendicular to the line connecting the carbons flanking it on either side. 
In addition, moves which regrow a whole segment in a configurationally biased way have been used.

\begin{figure}
   \includegraphics[width=0.45\textwidth]{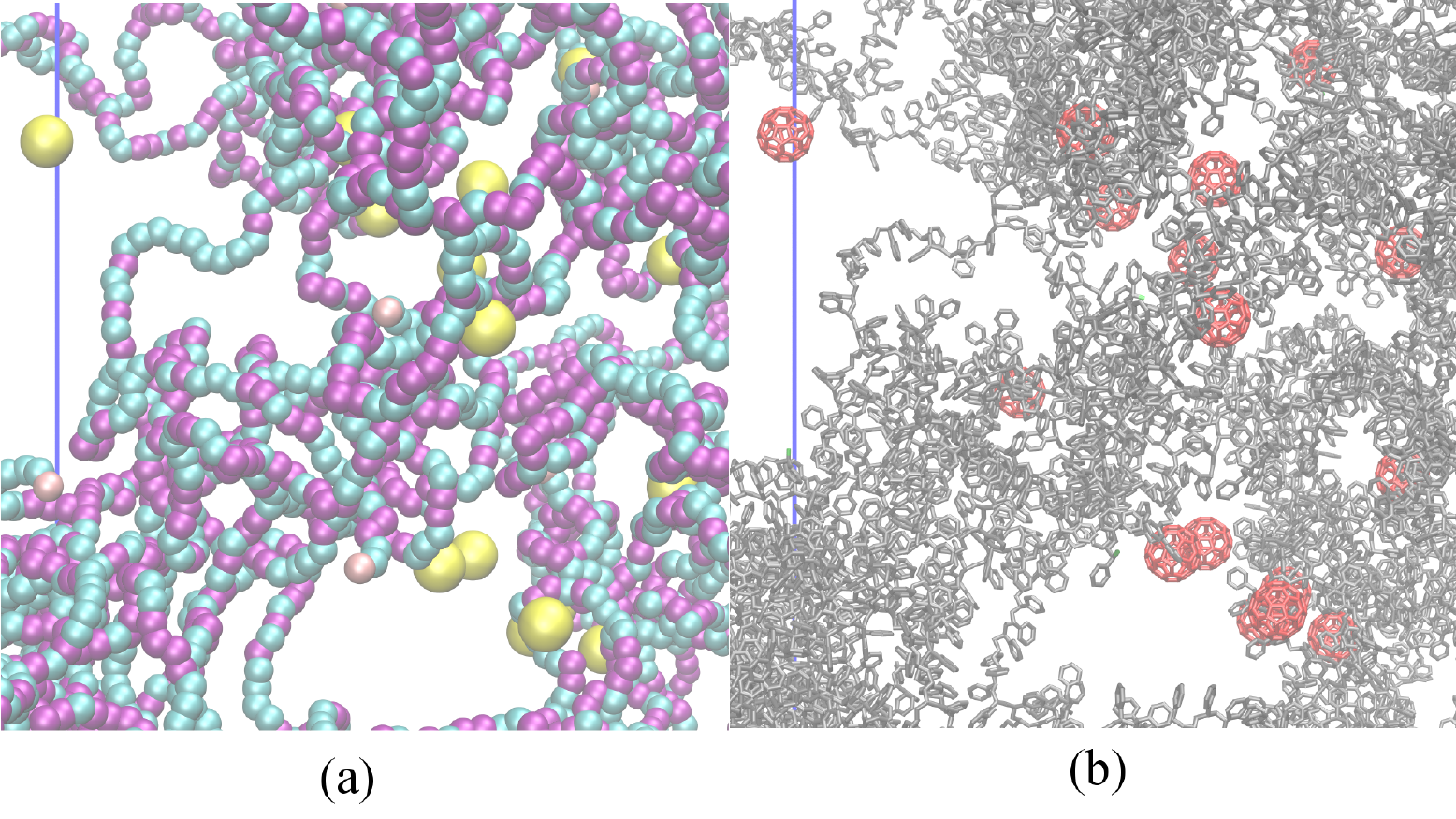}
   \caption{Illustration of the reverse mapping scheme. On the left hand side is shown a part of the simulation box 
   with fullerenes depicted as yellow spheres and PS chains as sequences of \textit{m} (cyan), \textit{r} (purple), and 
   \textit{em}/\textit{er} (brown) segments. On the right is the reverse-mapped atomistic system. 
   PS chains are unfolded and neighboring chains have been omitted for clarity. Visualization made
   by using the VMD software.\cite{JMolGraphics_14_33}}
   \label{fig:rm_schematic}
\end{figure}

The potential energy of the atomistic system is minimized with respect to the Cartesian coordinates of all atoms, 
excluding the \ce{CH2} united atoms which coincide with the positions of the coarse-grained sites. In this way, 
the equilibrated MC configuration is not distorted during the reconstruction of carbon atoms through reverse 
mapping ( \ref{fig:rm_schematic}).
For the minimization, LAMMPS is used with a Hessian-free truncated Newton minimization 
method.\cite{Nocedal_Wright_Numerical_Optimization} 
Minimization is performed in turns, by gradually blowing up the atomic radii.\cite{Macromolecules_18_1467}
One starts with atoms of reduced size (sigma equal to half its actual value), adjusting that size in stages so that 
the atoms reach their full size at
the end. As in earlier works, a modified potential energy function was used to describe non-bonded interactions in 
early stages of the minimization,
the so-called soft sphere potential. After the introduction of the Lennard-Jones interactions, the collision diameter, 
$\sigma$, is gradually increased from half to its final value. 
The reverse mapping scheme just described was designed in order to prevent locking of the local configuration in
torsional states which are inconsistent with the unperturbed conformational statistics adopted by PS in the melt, 
without departing at all from the well-equilibrated configurations provided by the coarse-grained simulations. 
Validation of the reverse-mapped configuration against experimentally available information from pure polystyrene melts
is necessary before proceeding to examine the properties of nanocomposites. It is discussed in the following section.

\subsection{Thermodynamic properties and structure of the reverse-mapped configurations}
The cohesive energy, $U_{\rm coh}$, is the energy associated with the intermolecular interactions only and can be
estimated by taking the difference between the total energy of the simulation box, $U_{\rm tot}$, and that of the 
isolated polymeric chains, $U_{\rm intra}$. In order to determine the intramolecular energy, chains were considered 
not to interact with their periodic images. Hildebrand's solubility parameter, $\delta$, is the square root of the 
cohesive energy density, $\delta = \left((U_{\rm intra} - U_{\rm tot})/V \right)^{1/2}$ with $V$ being the volume
of the simulation box.
The solubility parameter, $\delta$, was calculated for the bulk reverse mapped structures of 1460 diads per 
chain. For this calculation the intermolecular interactions for each system were summed, averaged and divided by
the simulation box volume to obtain an estimate of the cohesive energy density. 
The square root of the cohesive energy density, averaged over all structures, was found to be equal 
to $7 \:{\rm cal}^{1/2}{\rm cm}^{-3/2}$ at $T=500\:\text{K}$. 
Experimental values for PS from viscosity measurements in different solvents, 
range from $8.5$ to $9.3\:{\rm cal}^{1/2}{\rm cm}^{-3/2}$ at $298\:\text{K}$.\cite{TransFaradaySoc_54_1742}
The discrepancy of about 18\% is rather large and can be partly attributed to the temperature difference (the cohesion
of the polymer drops as the temperature rises and so $\delta$ should decrease as well). The short-chain structure
(made from the single 80-mer parent chain) equilibrated with MD at $500\:\text{K}$ gave an even lower solubility
parameter.

\begin{figure}
   \includegraphics[width=0.45\textwidth]{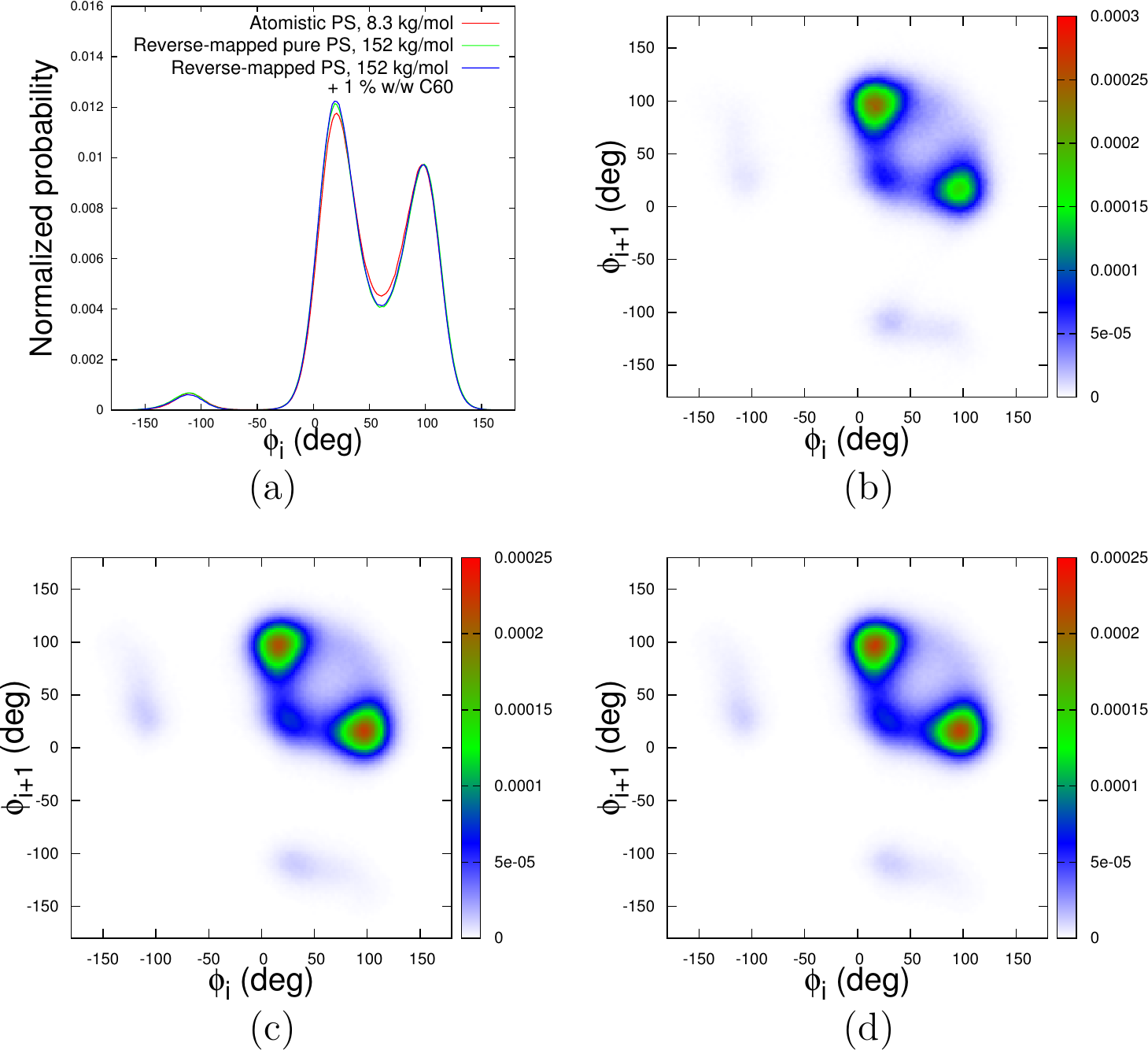}
   \caption{(a) Distribution of torsion angles for the \textit{meso} diads of 1460-mer chains 
            in the bulk systems obtained via reverse mapping (green line pure PS, blue line composite) and in 
            an atomistic 80-mer system directly equilibrated by MD (red line).
            (b,c,d) Ramachandran plots for pairs of successive torsion angles belonging to meso diads from the 
            80-mer atomistic system (b), bulk 1460-mer system (c) and the composite 1460-mer (d).}
   \label{fig:meso_torsion_distributions}
\end{figure}

\begin{figure}
   \includegraphics[width=0.45\textwidth]{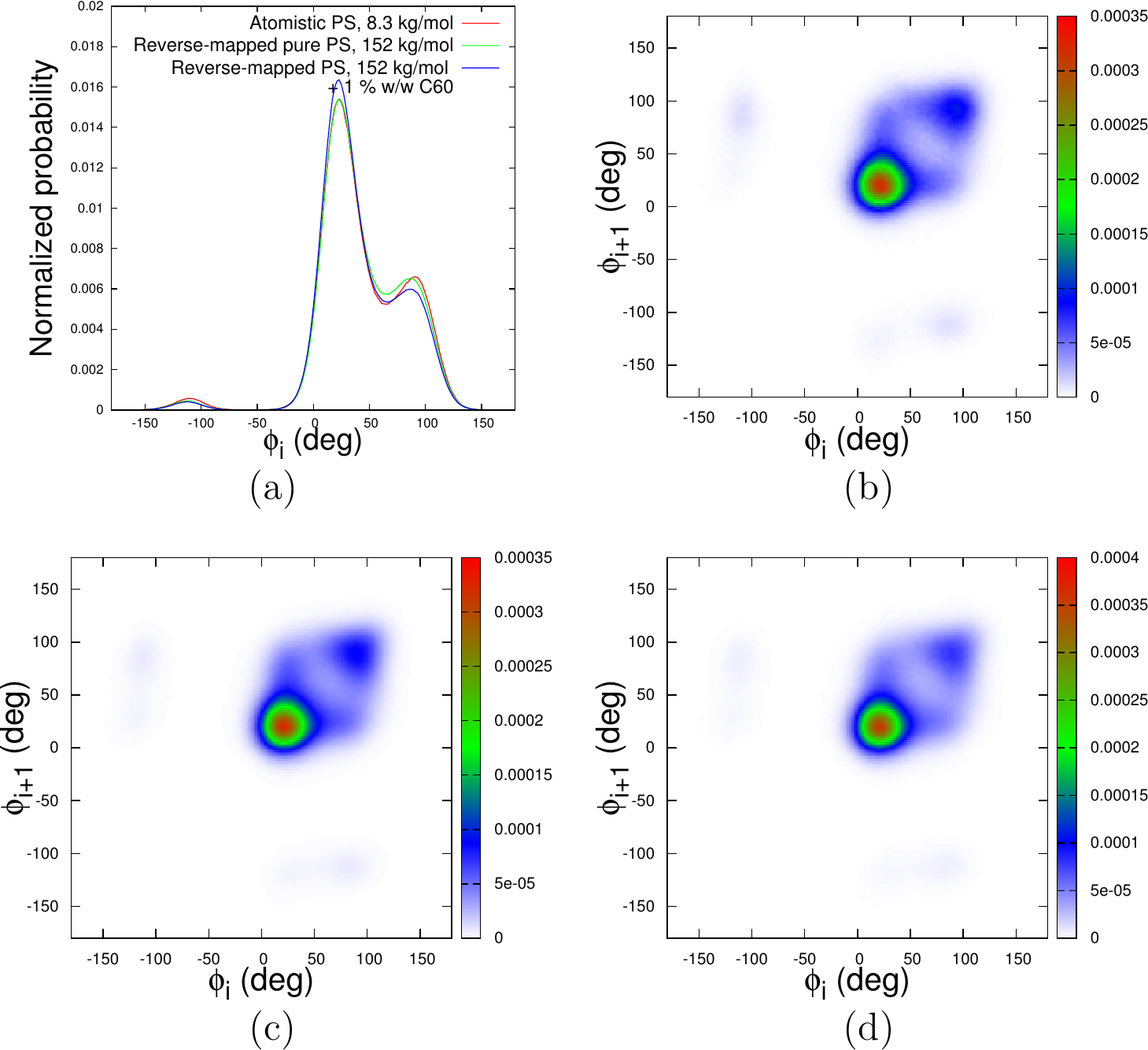}
   \caption{(a) Distribution of torsion angles for the \textit{racemo} diads of 1460-mer chains 
            in the bulk systems obtained via reverse mapping (green line pure PS, blue line composite) and in 
            an atomistic 80-mer system equilibrated directly by MD (red line).
            (b,c,d) Ramachandran plots for pairs of successive torsion angles belonging to meso diads from the 
            80-mer atomistic system (b), bulk 1460-mer system (c) and the composite 1460-mer (d).}
   \label{fig:racemo_torsion_distributions}
\end{figure}

NMR measurements on atactic PS have helped gain insight into the conformations adopted by its chains. Suter and 
collaborators \cite{Macromolecules_28_5320,Macromolecules_31_8918,Macromolecules_32_8681,MacromolTheorySimul_3_19}
have shown that considerable deviations may occur between experimental findings and predictions obtained from bulk
atomistic model structures of PS; furthermore, they have proposed several algorithms for the generation of atomistic
structures by an appropriate choice of the target conformational probabilities in the spirit of the Rotational Isomeric
State (RIS) model.
Having this in mind, the ability of the proposed reverse mapping scheme to generate atomistic configurations with 
correct conformational statistics is examined. 
The resulting torsional distributions are shown in Figures \ref{fig:meso_torsion_distributions} and 
\ref{fig:racemo_torsion_distributions} for meso and racemo diads, respectively.

\begin{table*}
   \caption{Torsion angle distribution, averaged all neat 1460mer reverse mapped structures}
   \label{tab:torsion_angle_distro}
   \begin{tabular}{cccc}
	   \hline
	   & \multicolumn{3}{c}{State frequency in \%} \\ \cline{2-4}
	   & \textit{t}: $-60^\circ \le \phi \le +60^\circ$	& \textit{g}: $+60^\circ < \phi \le +180^\circ$ 
	   & $\bar{g}$: $-180^\circ \le \phi <-60^\circ$ \\ \hline{}
	   \textit{meso} diads & 54.2 & 43.4 & 2.4 \\
	   \textit{racemo} diads & 66.7 & 31.7 & 1.6 \\
	   overall & 60.4 & 37.6 & 2.0 \\ \cline{2-4}
	   State a-priori RIS & & & \\
	   probability (300 K)\cite{MacromolTheorySimul_3_1} & 71 & 29 & - \\ \cline{2-4}
	   Double-quantum & & & \\
	   solid-state NMR\cite{SolidStateNuclMagnReson_12_119} & 68 & 32 & - \\ \hline
   \end{tabular}
\end{table*}

The fusion of \textit{trans} and \textit{gauche} states is apparently produced by the atomistic potential at this 
high temperature, since it is also present in the single-chain system which has been equilibrated by MD.
Comparing the curves in Figures \ref{fig:meso_torsion_distributions} and \ref{fig:racemo_torsion_distributions} we 
observe that torsion angle distributions in the reverse-mapped structures are extremely close to 
those obtained from the 80mer structure that was directly equilibrated by MD at the atomistic level, without intervention
of any coarse-graining and reverse mapping.
This observation is extremely promising for our reverse 
mapping scheme, implying that we can achieve well-equilibrated structures down to the atomistic scale. The percentage of 
\textit{trans} states is slightly increased in the presence of fullerenes, both for \textit{meso} and \textit{racemo} diads. 
In the case of \textit{racemo} diads, the effect is accompanied with an equal reduction of the \textit{gauche} 
percentage, which may be attributed to a mild extension of the chains trying to engulf the fullerenes. 
The overall content of trans conformations, as presented in \ref{tab:torsion_angle_distro} for the bulk systems,
in the reverse-mapped structures is around 60\%, which is in reasonable
agreement with the experimental $68 \pm 10 \:\text{\%}$ measured by NMR.\cite{SolidStateNuclMagnReson_12_119}
The total percentage of $\bar{\rm g}$ conformations in the reverse-mapped structures was 1.8\%, while, according to 
RIS models, it should be less than 2\%.\cite{JAmChemSoc_91_3111,Macromolecules_3_43,MacromolTheorySimul_3_1}
Clearly, capturing the correct torsion angle distribution is a stringent test for reverse mapping from the 
coarse-grained representation adopted in this work. Previous efforts by Spyriouni et al.,\cite{Macromolecules_40_3876}
and Ghanbari et al. \cite{Macromolecules_44_5520} could not capture the correct local structure of polystyrene, 
leading to a high percentage of unrealistic $\bar{\rm g}$ conformations.

Along with the torsion angle distributions in Figures \ref{fig:meso_torsion_distributions} and 
\ref{fig:racemo_torsion_distributions}, Ramachandran plots of the two-dimensional distributions characterizing two 
successive torsion angles are also shown. 
Following Flory et al.,\cite{JAmChemSoc_96_5015} in the case of a \textit{meso} diad, $\phi_i$ is measured in the 
right-handed sense and $\phi_{i+1}$ in the left-handed sense. 
In the \textit{racemo} diad obtained by inverting the chirality of the second
methine carbon of the diad, both torsion angles are measured in the right-handed sense (the mirror image diad would 
require two left-handed frames of reference).
Overall, the convention used to define the sense of rotation of the torsional angles is such that the same angles 
lead to the same molecular environments around the considered bond.\cite{Macromolecules_8_776}
In  \ref{fig:meso_torsion_distributions}(b) the plot concerns the single-chain 80-mer system. 
It can be seen that for this system, \textit{tg} and \textit{gt} conformation probabilities are not equal, since the 
two regions are not evenly populated; this reflects incomplete equilibration by MD, even in this short-chain melt. 
On the other hand, systems produced by the proposed reverse mapping methodology, 
result in fully symmetric \textit{tg} and \textit{gt} conformations.
The presence of fullerenes does not seem to affect the probabilities of successive torsion angles.
In all cases, conformations involving \textit{\=g} torsion angles are extremely rare.

\section{Atomistic Molecular Dynamics}
All MD simulations have been conducted using the Large-scale Atomic/Molecular Massively Parallel Simulator 
(LAMMPS),\cite{JCompPhys_117_1} extended with the united-atom force field of Lyulin and 
Michels,\cite{Macromolecules_35_6332} which we have incorporated into the LAMMPS source code.
The equations of motion are those of Shinoda et al.,\cite{PhysRevB_69_134103} integrated by the time-reversible 
measure-preserving Verlet integrator derived by Tuckerman et al.\cite{JPhysAMathCen_39_5629}
In all cases, a timestep of 1 fs was used. 

Initially, the reverse-mapped configurations were subjected to 50 ns of isothermal-isobaric ($NpT$) MD under 
$T=500\;{\rm K}$ and $p=101.325\;{\rm kPa}$, using the barostat of LAMMPS. Keeping the temperature fixed, 20 ns of 
isothermal ($NVT$) integration followed, leading to the final 100 ns time integration under constant energy ($NVE$), 
where the coordinates of the atoms were tracked in order to extract the dynamical properties. 
The final configuration from the melt at $500\;{\rm K}$ was subjected to further $NpT$ simulation with the set 
temperature $T$ lowered by 20 K every 40 ns (effective cooling rate 0.5 K/ns) down to a final temperature of 380 K. 
At every cooling step (20 K), a configuration of the system was recorded and used for 50 ns of $NpT$ equilibration, 
followed by 20 ns $NVT$ and 100 ns $NVE$ MD run in order to extract the dynamics at this temperature.
All (three neat and three composite) independent reverse-mapped configurations were subjected to the same procedure. 
During the $NVE$ run, the system's pressure and temperature were monitored in order to ensure that they correspond to 
the desired values.

We believe that $NVE$ simulations, where no barostatting or thermostatting take place, are the best means of studying 
dynamics under no external influence.\cite{ComputerSimulationOfLiquids} 
The thermostatting and barostatting is achieved by adding some dynamical variables which are coupled to the particle 
velocities (thermostatting) and simulation domain dimensions (barostatting), in order to mimic a reservoir coupled to 
the system. If the coupling is loose, the energy flow from the system to the 
reservoir is slow. On the other hand, if the coupling is strong, long-lived weakly damped oscillations in the energy
occur resulting in poor equilibration. It is necessary to choose the strength of the coupling wisely,
so as to achieve satisfactory damping of these correlations.\cite{MolPhys_52_255} 

\subsection{Hydrogen reconstruction}
Hydrogen reconstruction aims at re-introducing hydrogens of \ce{CH3}, \ce{CH2} groups and phenyl rings. The procedure 
we followed is inspired by the work of Ahumada et al.\cite{Macromolecules_35_7110}
Methyl hydrogens are reconstructed at a staggered conformation,
i.e. a methyl C-H bond being coplanar with the methyl stem and with the C-H bond of the methine group to which the 
methyl is connected and pointing in an opposite direction to the latter bond. 
For all methyls, C-H bonds are assumed to be $b_{\rm C-H}= 1.10\;\text{\AA}$ long and to form an angle 
of $\theta_{\rm C-C-H}=110^\circ$ with the 
methyl stem.
As far as the \ce{CH2} united atoms are concerned, two hydrogen atoms are placed symmetrically on both sides of the 
plane where the C-C bond lies, obeying the equilibrium bond length $b_{\rm C-H}$ and equilibrium bond angle 
$\theta_{\rm C-C-H}$. 
In the case of the \ce{CH} united atom of the backbone, the single hydrogen atom is placed entirely symmetrically to 
the phenyl ring, using as a plane of symmetry the backbone of the chain.
Finally, one hydrogen atom is attached to every aromatic atom in the direction defined by the center of mass of the 
ring and the carbon atom, at a distance $b_{\rm C-H}$ from it. The addition of hydrogens to an end of a polystyrene 
chain is depicted in Figure S1 of the Supporting Information to the present paper.

\subsection{Temperature dependence of segmental dynamics}
We examine the segmental dynamics of the atactic PS melt, as predicted by the united-atom MD simulations, by 
analyzing time autocorrelation functions of various vectors. In the case of polystyrene, the vectors characterizing 
the orientation of the phenyl ring and the orientation of the C-H bonds are of special interest
(Figure S1 of the Supporting Information to the present paper).
The orientational decorrelation with time for each one of these vectors can be studied by considering 
ensemble-averaged Legendre polynomial of order $k$, $P_{k}(t)$, of the inner product 
$\left\langle \mathbf{v}(t_0) \cdot \mathbf{v}(t_0 +\Delta t)\right \rangle$ of the unit vector
$\mathbf{v}$ along the vector, at times $t_0$ and $t_0+t$. 
It is quite common to fit the long-time behavior of orientational autocorrelation functions of this kind by a modified Kohlrausch - 
Williams - Watts (mKWW) function \cite{AnnPhys_167_1,TransFaradaySoc_66_80,Macromolecules_40_2235}:
\begin{align}
P_{k} (t) = & \alpha_\text{lib} \exp{\left[-\frac{t}{\tau_\text{lib}}\right]} \nonumber \\
& + \left(1- \alpha_\text{lib}\right) \exp{\left[-\left(\frac{t}{\tau_\text{seg}} \right)^{\beta_\text{KWW}}\right]} 
\label{eq:mKWW}
\end{align}
The mKWW function of eq \ref{eq:mKWW} consists of two parts. The first term describes a fast exponential decay with amplitude
$\alpha_{\rm lib}$, which is associated with the fast librations of torsion angles around skeletal
bonds and with the bond stretching and bond angle bending vibrations of skeletal and pendant bonds near their equilibrium values, 
with characteristic time $\tau_{\rm lib}$. The second term is a stretched exponential decay associated with 
cooperative conformational transitions in the polymer, with $\tau_{\rm seg}$ being the characteristic correlation time 
and $\beta_{\rm KWW}$ the stretching exponent.
The correlation time for segmental motion, $\tau_{\rm c}$, also referred to as ``segmental relaxation time'' in the 
following, can be calculated as the integral:
\begin{align}
\tau_\text{c} & = \int_0^\infty P_k(t) dt \nonumber  \\ 
& =  \alpha_\text{lib} \tau_\text{lib}
 + (1 - \alpha_\text{lib}) \tau_\text{seg}
\frac{1}{\beta_\text{KWW}} \Gamma\left(\frac{1}{\beta_\text{KWW}}  \right)
\label{eq:tcor_mKWW}
\end{align}

\begin{figure}
   \includegraphics[width=0.45\textwidth]{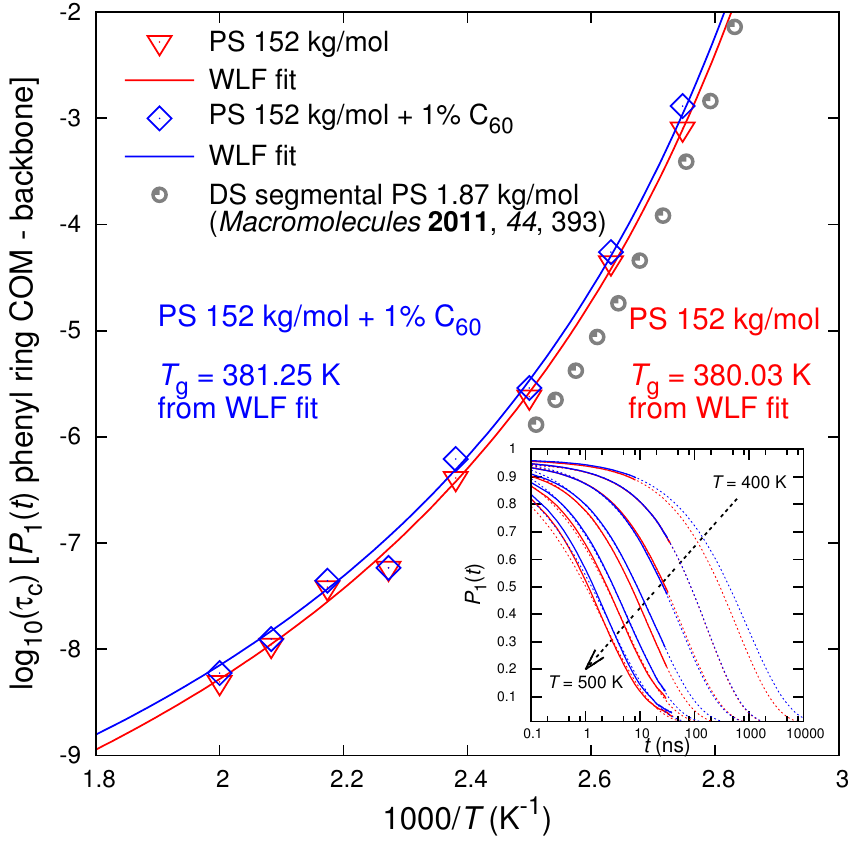}
   \caption{Temperature dependence of the relaxation times obtained from MD simulations corresponding to the $P_1(t)$
   autocorrelation function of the vector connecting the backbone CH group with the center of mass of the phenyl ring. 
   Experimental points come from DS measurements on neat low-molar mass PS.\cite{Macromolecules_44_393} Fits to the 
   WLF\cite{JAmChemSoc_77_3701} equation are also presented. 
   In the inset to the figure, $P_1(t)$ time autocorrelation
   functions from MD simulations (solid lines) are depicted, along with their fits to eq \ref{eq:mKWW} (dashed lines).}
   \label{fig:p1_ring_com_wlf}
\end{figure}

If a comparison with dielectric spectroscopy (DS) is sought, the relevant vector to study is the vector starting from 
the backbone \ce{CH} united atom and ending at the center of mass of the phenyl ring, $\mathbf{v}_{\rm CH-CM}$. 
To a good approximation, monomer dipole moments are directed along this vector.
DS measurements cannot discern between self and cross correlations of the dipole vectors. 
However, the contribution of the correlations of neighboring dipole moments to the  segmental relaxation is minimal.
\cite{JChemPhys_106_3798,JChemPhys_119_6883}
In this case, the quantity of interest is the Legendre polynomial of the first kind, 
$P_1(t) = \left\langle \hat{\mathbf{v}}_{\rm CH-CM}(t_0 + t) \cdot \hat{\mathbf{v}}_{\rm CH-CM}(t_0) \right\rangle$.
In the inset to the  \ref{fig:p1_ring_com_wlf} the calculated $P_1(t)$ functions are presented with solid lines
for both the bulk and the composite systems. Along with the simulation results, fits to the mKKW function 
(eq \ref{eq:mKWW}) are also presented. It can be seen that the mKWW expression describes well the simulation results for 
both cases and the whole temperature range. Each simulation curve represents the average of the three independent MD 
trajectories produced by the different reverse-mapped structures. 
The fitting to the mKWW equation allows us to analytically estimate the segmental 
relaxation time, $\tau_{\rm c}$, based on eq \ref{eq:tcor_mKWW}, using the fit parameters of 
Table S1 of the Supporting Information to the present paper.

The temperature dependence of the segmental relaxation times is presented in the main part of  
\ref{fig:p1_ring_com_wlf}. Segmental relaxation times calculated by our 
united-atom MD simulations are in favorable agreement with experimental DS measurements found in the work of 
Harmandaris et al.\cite{Macromolecules_44_393} Experimental points from ref \citenum{Macromolecules_44_393} are 
shifted to smaller timescales, due to the 
smaller molecular weight of the samples used in the DS measurements. This is expected, since PS dynamics exhibits
molecular weight dependence which can shift the glass transition temperature from $314\;{\rm K}$ for 
$1.35\;{\rm kg/mol}$ PS to $373.3\;{\rm K}$ for $243\;{\rm kg/mol}$ PS.\cite{Macromolecules_41_9335}
An estimate of the glass transition temperature, $T_{\rm g,sim}$ can be obtained by fitting the temperature 
dependence of segmental relaxation times to an equation such as Williams-Landel-Ferry (WLF).\cite{JAmChemSoc_77_3701}
Fitted values of $T_{\rm g}$ are given in  \ref{fig:p1_ring_com_wlf}, with the coefficient of WLF equation,
$\log{\left(\tau_{\rm c}/ \tau_{\rm c,g}\right)} = 
- \left[c_1 \left(T - T_{\rm g} \right) \right]/\left(T - T_{\rm g} + c_2 \right)$,
being $c_1 = 13.6$ and $c_2 = 56\;{\rm K}$, close to the universal values and the experimental ones of Kumar et al.
\cite{JChemPhys_105_3777}
As can be seen in  \ref{fig:p1_ring_com_wlf}, nanocomposite systems exhibit slightly longer segmental 
relaxation times, compared to their neat 
counterparts, for the majority of temperatures studied. 
This leads to an estimated glass transition shift of around $1\;{\rm K}$ upon the addition of fullerenes. 
The $T_{\rm g}$-shift predicted by the MD simulations is in excellent agreement to the shift found by differential 
scanning calorimetry (DSC) measurements of Kropka et al.\cite{NanoLett_8_1061}
However, for some temperatures, the dynamical behavior of unfilled and filled systems yields completely 
indistinguishable results. 

If one is interested in comparing with NMR data, an appropriate autocorrelation function to look at is the orientational
autocorrelation function of C-H bonds. 
The reason is that, for $^2{\rm H}$ nuclei, spin-lattice relaxation is dominated by electric quadrupole coupling and the
spin relaxation time can be directly related to the reorientation of the C-$^2{\rm H}$ bond. In this case, the second 
Legendre polynomial of the unit vector $\hat{\mathbf{v}}_{\rm b}$ directed parallel to a C-H bond:
$P_\text{2}(t) = \frac{3}{2} \left \langle \left(\cos{\theta_{\rm b}(t_0,t)}\right)^2 \right\rangle -\frac{1}{2}
= \frac{3}{2} \left \langle \left(\hat{\mathbf{v}}_{\rm b}(t_0+t)\cdot \hat{\mathbf{v}}_{\rm b}(t_0)  
\right)^2 \right\rangle -\frac{1}{2}$
is employed as a measure of the polymer segmental dynamics. $\theta_{\rm b}(t_0,t)$ is the angle of the bond vector 
$\mathbf{v}_{\rm b}$ at time $t$ relative to its original position at the time origin $t_0$ and the brackets 
$\left\langle ... \right\rangle$ denote an ensemble average over all C-H bonds in the system and across different time 
origins.
Since our MD simulations were conducted using a united-atom model, hydrogens were reconstructed upon post-processing
the trajectories, following the procedure described above. 

\begin{figure}
   \includegraphics[width=0.45\textwidth]{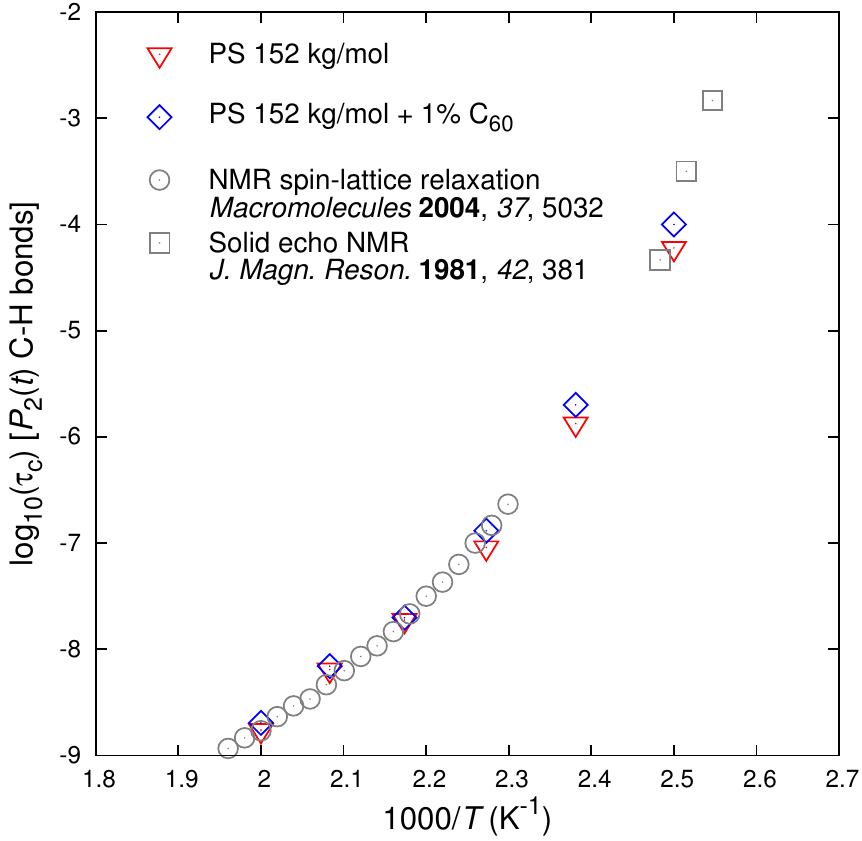}
   \caption{Temperature dependence of the segmental relaxation times obtained through analysis of the $P_2(t)$ curve 
   of the C-H bond vectors. Experimental points correspond to spin-lattice 
   relaxation\cite{Macromolecules_37_5032} and solid echo\cite{JMagnReson_42_381} NMR measurements.}
   \label{fig:p2_aliphatic_wlf}
\end{figure}

 \ref{fig:p2_aliphatic_wlf} presents the segmental correlation times, as extracted from the C-H vector $P_2(t)$ 
autocorrelation functions from the simulation trajectories. The best fit parameters for the mKWW equation used are 
reported in Table S2 of the Supporting Information to the present paper.
To be consistent with the NMR measurements, the weighted average autocorrelation function over the eight C-H vectors 
of each monomer was taken into account for the estimation of relaxation times. Again, $P_2(t)$ functions were fitted 
with a mKWW equation (eq \ref{eq:mKWW}) in order to predict the relaxation time. 
C-H bond reorientation relaxation times are found to be in good agreement with the experimental data of 
He et al.\cite{Macromolecules_37_5032} obtained by NMR spin-lattice relaxation experiments and with the solid echo NMR 
measurements of Spiess and Sillescu.\cite{JMagnReson_42_381}
Despite the fact that our united-atom model with reconstruction of hydrogens exhibits faster dynamics at short time 
scales, it can capture reasonably well the evolution of autocorrelation functions at long timescales.
Throughout the temperature range studied in our MD simulations, the segmental relaxation times are found to be in 
excellent agreement with the experimental measurements. Moreover, the nanocomposite system under study exhibits slower
segmental dynamics than the bulk. However, as in the case of $P_1(t)$ analysis, the overall dynamics of the systems
are really close to each other, prompting the need for a local analysis of the segmental dynamics.

\subsection{Many-nanoparticle influence on dynamics} 
\begin{figure}
   \includegraphics[width=0.45\textwidth]{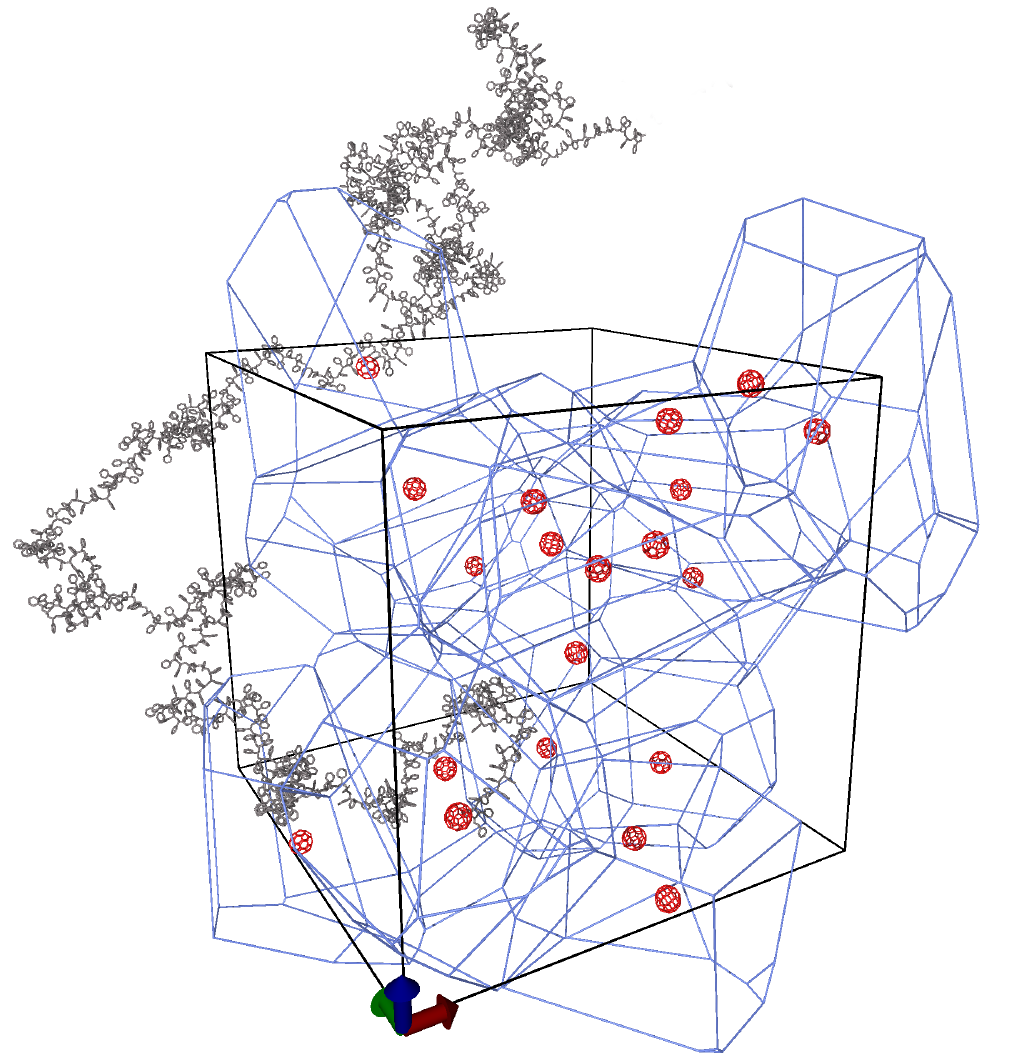}
   \caption{Schematic of the Voronoi tessellation of the simulation box. One unfolded atomistic 1460mer chain is 
   also shown. Dispersed fullerenes serve as the centers of the Voronoi cells.}
   \label{fig:voronoi_schematic}
\end{figure}

The study of local dynamics, when many nanoparticles are present, requires a tessellation of space, so as to 
quantify possible many-body effects. Two possible tessellation schemes are the partitioning of the space taken up by 
the material into Voronoi polyhedra, or their duals, Delaunay tetrahedra.
Inspired by the work of Starr et al.\cite{PhysRevLett_89_125501} on glass-forming liquids, we choose to carry out an 
analysis based on the Voronoi tessellation of the simulation box with fullerenes acting as the centers of the 
Voronoi cells ( \ref{fig:voronoi_schematic}).
Voronoi polyhedra provide a direct way of quantifying the confinement imposed by a fullerene on its neighborhood. 
Smaller distances between neighboring fullerenes yield Voronoi cells of smaller volume. Thus, from now on, we will 
employ the volume of a Voronoi cell as a measure of the confinement experienced by the polymer lying in it (the 
smaller the cell the more confined the polymeric matrix around the specific fullerene). 
Some convenient features of the Voronoi tessellation, over its dual, are the constant number of cells (which is equal
to the number of dispersed nanoparticles) and the significantly larger volume of the cells (since the Voronoi cells
are always by one order of magnitude fewer than the Delaunay tetrahedra). 
The partitioning of the simulation box is carried out by using the well-established Voro++ software library of 
Rycroft et al.\cite{Chaos_19_041111, PhysRevE_74_021306}

\subsection{Local mean-squared displacement of backbone carbon atoms}
A rigorous way of studying the mobility of a polymeric melt is to calculate the mean-squared displacement (MSD) of 
backbone carbon atoms. 
In order to avoid chain end effects,\cite{JChemPhys_116_436,Macromolecules_36_1376} only the innermost backbone 
carbons along the chain contribute to the calculations:
\begin{equation}
g(t) = \frac{1}{2n_{\rm inner}+1}\sum_{i=N/2-n_{\rm inner}}^{N/2+n_{\rm inner}}
\left \langle \left(\mathbf{R}_i (t_0 + t) - \mathbf{R}_i(t_0)\right)^2 \right \rangle
\end{equation}
with the value of the parameter $n_{\rm inner}$ quantifying the number of innermost atoms, on each side of the middle 
segment of each chain that are monitored. In our case, $n_{\rm inner}$ is set in such a way that we track half of the chain, 
excluding one fourth of the chain close to one end and one fourth close to the other end. 

\begin{figure}
   \includegraphics[width=0.45\textwidth]{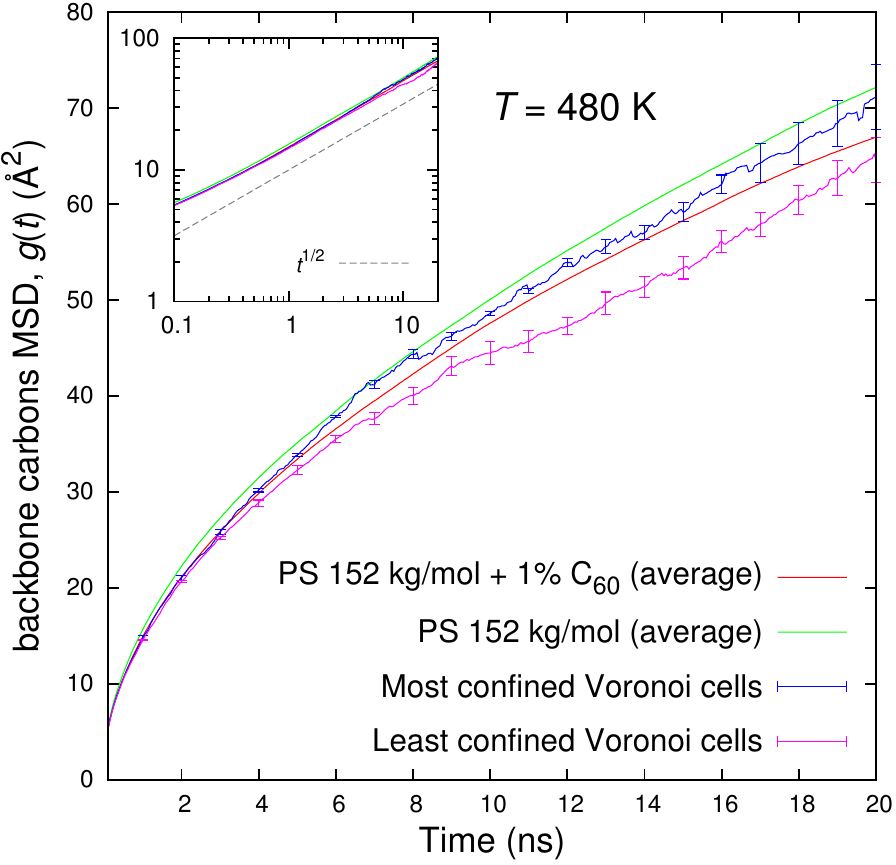}
   \caption{Mean-squared atomic displacements of backbone carbon atoms as a function of time for filled and unfilled 
   polystyrene systems at $T=400\;{\rm K}$. 
   In the case of fullerene nanocomposites, an analysis of the dependence of backbone MSD on 
   confinement is also presented for most and least confined Voronoi cells (indicative error bars also included). 
   In the inset to the figure, the same data are presented in logarithmic axes.}
   \label{fig:local_msd_480}
\end{figure}

 \ref{fig:local_msd_480} presents the mean-squared displacement of backbone carbon atoms as a function of time 
at a temperature of $480\;{\rm K}$ for both the filled and unfilled systems. 
As can be seen, nanocomposite systems exhibit lower mobility when compared to their neat counterparts. 
The MSD of backbone carbons is depressed upon the 
addition of fullerenes, in good agreement with the neutron scattering observations of Kropka et al.\cite{NanoLett_8_1061}
In the inset to  \ref{fig:local_msd_480}, a logarithmic plot of the functions $g(t)$ is presented. 
The scaling of $g(t) \sim t^{1/2}$ is expected for the very short time behavior 
studied.\cite{Doi_Edwards_TheoryOfPolymerDynamics} As the Rouse model predicts, the segments do not feel the 
constraints of the entanglement network around them, following a Brownian motion in the free space available. 
This behavior is expected for times $t$ shorter than the characteristic time $\tau_{\rm e}$ ($t\le \tau_{\rm e}$) 
when the segmental displacement becomes comparable to the tube diameter. 
Likhtman and McLeish \cite{Macromolecules_35_6332} estimated that the 
time marking the onset of the effect of topological constraints on segmental motion, $\tau_{\rm e}$ is 
$3.36\cdot10^{-4}\;{\rm s}$ for polystyrene. Since the results presented in  \ref{fig:local_msd_480} go up 
to $20\;{\rm ns}$, the scaling of $t^{1/2}$ is fully justifiable.

We now move to the estimation of the local MSD, for the timespan an atom spends inside a particular cell of the Voronoi
tessellation. In our analysis we use the average MSD from the three most confined and three least confined cells, 
averaged over the three independent configurations created. 
We have observed that the volume of the Voronoi cells does not change significantly as a function of time. 
Based on this analysis for the nanocomposite system, the degree of depression is found to be a function of the 
confinement induced by the fullerenes. The diffusion of chains is spatially inhomogeneous, as observed by 
Desai et al.\cite{JChemPhys_122_134910}
Small Voronoi cells tend to lead to higher mobility of the segments. This suggests an image of fullerenes
as small grinders dispersed in the polystyrene matrix. Despite the fact that the addition of fullerenes limits the 
diffusion of polymeric chains, there exist regions in space, where the polymer can recover part of its dynamics due to 
the high level of confinement. 

\begin{figure}
   \includegraphics[width=0.45\textwidth]{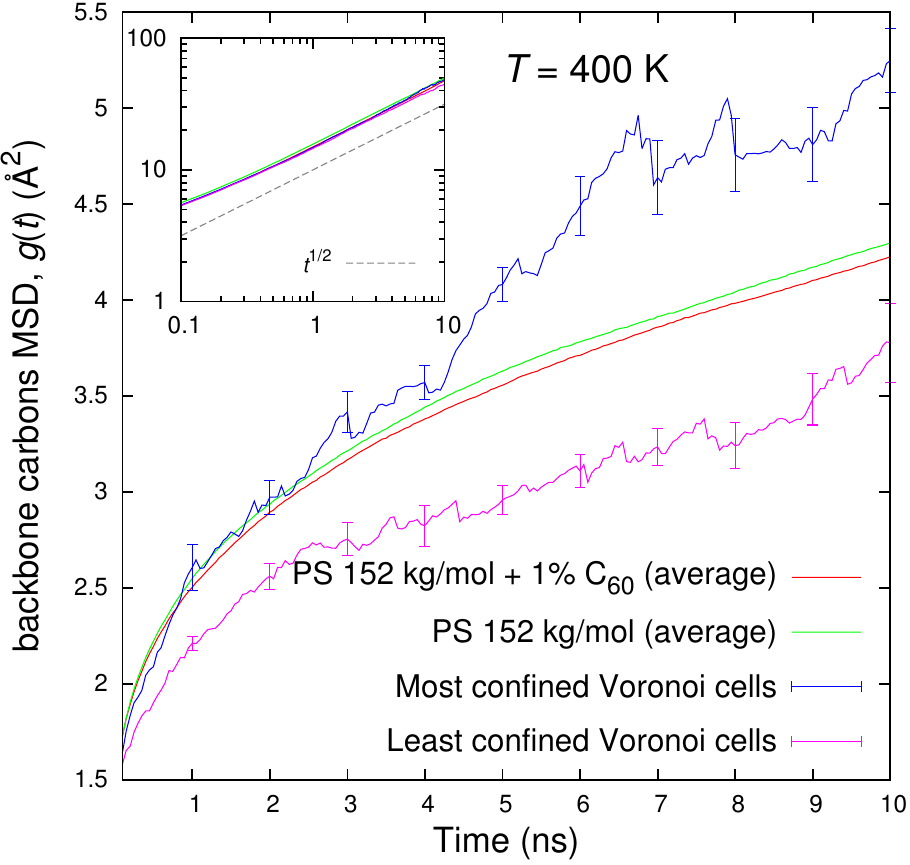}
   \caption{Mean-squared atomic displacements of backbone carbon atoms as a function of time for filled and unfilled 
   polystyrene systems at $T=400\;{\rm K}$. 
   In the case of fullerene nanocomposites, an analysis of the dependence of backbone MSD on 
   confinement is also presented for the most and least confined Voronoi cells (indicative error bars also included). 
   In the inset to the figure, the same data are presented in logarithmic axes.}
   \label{fig:local_msd_400}
\end{figure}

The dynamic heterogeneity of the nanocomposite systems grows as the temperature is lowered. The local mean-squared 
displacement of the backbone atoms, as a function of time, at a temperature of $400\;{\rm K}$ is presented in 
 \ref{fig:local_msd_400}. As happens at higher temperature, the average MSD of the nanocomposite system is 
smaller compared to the average MSD of the bulk one. However, carbon atoms lying in the most confined Voronoi cells 
present larger displacements than the ones lying in the least confined. It can be observed that, despite the fact that MSD 
absolute values are getting smaller as expected, their variance when one studies local dynamics is larger.  
This observation strengthens our hypothesis that fullerenes act as nanoscopic grinders dispersed in the polymeric matrix, 
yielding strong deviation of polymer local dynamics from the bulk in their close neighborhood.

\subsection{Fullerene rotational diffusivity}
The fact that the fullerenes are geometrically rigid on an atomistic scale (although clearly vibrating) and essentially 
spherical enables their rotational motion to be readily characterized using a single rotational diffusion measure.
If $\mathbf{e}$ is any arbitrary unit vector embedded in the cluster and passing through its center, 
then the orientational correlation function $C_{\rm e}(t)=\left \langle \mathbf{e}_i(t) \mathbf{e}_i(0) \right \rangle$
can be computed. The simplest approach to define $\mathbf{e}$ is to use the separation vector between each 
fullerene's center of mass and a specific predefined atom.  

\begin{figure}
   \includegraphics[width=0.45\textwidth]{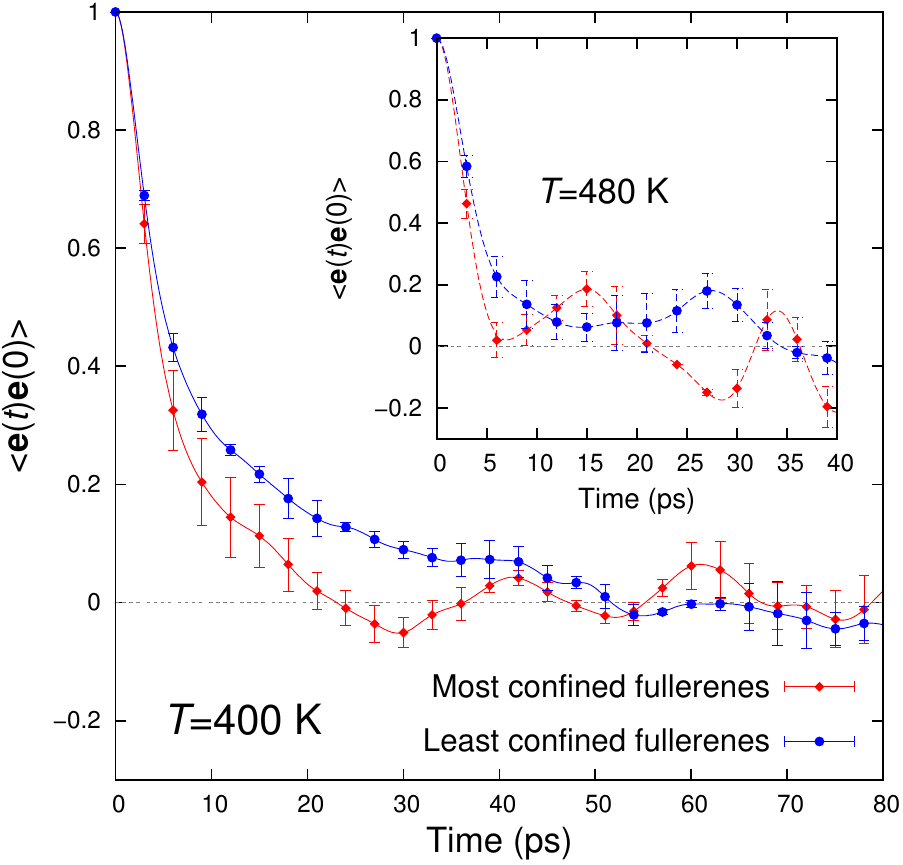}
   \caption{Fullerenes' rotational autocorrelation function. We consider a vector connecting a fullerene's 
   center of mass with a predefined atom. The main figure refers to 400 K, while the inset to the figure 
   refers to 480 K.}
   \label{fig:fullerene_rotdec}
\end{figure}

The results for two temperatures (which were considered for the local MSD calculation), are presented in  
\ref{fig:fullerene_rotdec}. The main plot contains the results for the temperature of 400 K, while in the inset the 
rotational decorrelation function at 480 K is plotted. Despite the noise in the measurements (since only the three most
confined and three least confined fullerenes contribute), it is clear that fullerenes which lie in the most confined 
cells rotate faster. There are many signs of anomalous rotational diffusion of fullerenes, which imply a strongly 
heterogeneous environment. An in-depth study of the fullerenes' rotational diffusion is outside the scope of the present 
work. However, this surprising observation coheres to the image of the nanoparticles acting as nanoscopic grinders
which force the polymeric chains to translate in their immediate neighborhood.

\section{Local stresses}
\subsection{Distributions of atomic-level stresses}
Atomic-level stresses can serve as a basis for characterizing local structure. Egami et al. \cite{PhilosMagA_41_883}
first applied atomic-level stresses to glasses in an atomistic computer model of amorphous iron. 
Theodorou and Suter \cite{Macromolecules_19_379} were the first to apply the idea of atomic level stresses to 
polymeric glasses, where both bonded and non-bonded interactions contribute to the stress. 
Following ref \citenum{Macromolecules_19_379}, we 
define the atomic stress tensor for atom $i$, in a system with central forces by
\begin{equation}
\sigma_{i, LM} = -\frac{1}{V_i}m_i v_{i,\rm L} v_{i,\rm M}
-\frac{1}{2V_i} \sum_{j \ne i} \left(r_{i,L} - r_{j,L} \right)^{\rm min.im.} \: F_{ij,\:M}^{\rm min.im.}
\label{eq:atomic_stress_definition}
\end{equation}
where $\mathbf{r}_{i}$ and $\mathbf{r}_{j}$ are the position vectors of atoms $i$ and $j$, 
$\mathbf{v}_i$ the velocity of atom $i$ and $\mathbf{F}_{ij}$ is the 
force exerted on atom $i$ by atom $j$.
The indices $L$ and $M$, indicating the three coordinate directions in a Cartesian system, assume the values $x$, $y$ 
and $z$. The superscript ``min.im.'' indicates interatomic distances and forces calculated according to the ``minimum
image convention''.
Thompson et al.\cite{JChemPhys_131_154107} have described 
different ways of formulating per-atom and global virial and stress calculations, including how it is done in  
LAMMPS.\cite{JCompPhys_117_1}
According to their formulation, a virial contribution produced by a small set of atoms (e.g. 4 atoms in a dihedral
angle) is assigned in equal portions to each one of these atoms.

In order to convert the virial into a stress, a local volume $V_i$ has to be associated with each atom.
Here, we shall use a Voronoi tessellation to define atomic volumes, \cite{Chaos_19_041111, PhysRevE_74_021306}
such that summing over all atoms gives the total volume of the system, $V = \sum_i V_i$.
Note that this atom-level Voronoi tessellation is much finer than the Voronoi tessellation with respect to fullerene 
centers used to partition the sample volume among different fullerenes.
Summation of all atomic-level stresses $\sigma_i$, multiplied by the appropriate volumes
$V_i$, yields the macroscopic stress; for a system in detailed mechanical equilibrium the ``internal'' stress tensor
$\sigma$ is recovered (this is a restatement of the virial theorem):\cite{MacromolTheorySimul_2_191}
\begin{equation}
\sigma_{\rm LM} = \frac{1}{\sum_i V_i} \sum_i V_i \sigma_{i,\rm LM}
\label{eq:macro_stress}
\end{equation}
Note that, as defined in eq \ref{eq:atomic_stress_definition}, per-atom stress is the negative of the per-atom pressure
tensor divided by an appropriate atomic volume. 
Thus, if the diagonal components of the per-atom stress tensor are summed for all atoms in the system and
the sum is divided by $3V$ where $V$ is the volume of the system, the ensemble average of the result should be $-p$, 
with $p$ being the total pressure of the system. 

We introduce two invariants of the atomic-stress tensor\cite{PhilosMagA_41_883,Macromolecules_19_379}
for characterizing local structure. The first is the ``atomic-level hydrostatic pressure'' $p_{i}$, defined by: 
\begin{equation}
p_{i} = \frac{1}{3} {\rm Tr}{\left({\boldsymbol \sigma_{i}}\right)}
\label{eq:atomic_p_definition}
\end{equation}
Although $p_{i}$ is termed a ``pressure'', it is 
really a tension or negative pressure. It can serve as a measure of local density fluctuations in the material. 
A high, positive value of $p_{i}$ is associated with a high coordination number around atom $i$ and with a 
lower than average atomic density. A low, negative value of $p_{i}$ is associated with a low coordination number and 
with higher than average local atomic density. 
The second invariant is the ``atomic-level von Mises shear stress'', $\tau_i$, defined by
\begin{equation}
\tau_i^2 = \frac{1}{2} {\rm Tr}\left[\left(\boldsymbol \sigma_{i} - p_{i}\mathbf{I}\right)^2\right]
\label{eq:atomic_von_Mises}
\end{equation}
where $\mathbf{I}$ is the matrix representation of the unit tensor. The quantity $\tau_i$ reflects the degree of 
asymmetry of the local environment around atom $i$.


Atomic-level stresses were calculated for an ensemble of 1200 configurations obtained during $NVE$ MD simulation at a 
temperature of $420\;{\rm K}$.
The characteristic quantities $p_i$ and $\tau_i$ were calculated according to eq \ref{eq:atomic_p_definition} and 
eq \ref{eq:atomic_von_Mises}, respectively. The distributions of $p$ and $\tau$ were accumulated seperately for each 
type of united atom interaction site (aliphatic \ce{CH} and \ce{CH2}, aromatic C and \ce{CH}, fullerene C) present in 
our system.
These distributions are plotted in Figures S2 and S3 of the Supporting Information to the present paper.
A striking feature, evident from the above figures, is the expected disparity in magnitude between the atomic-level
stresses $\sigma_i$ and the box (macroscopic) internal stress $\sigma$.\cite{Macromolecules_19_379} There is a strong 
compensation effect in the summation of atomic-level stresses to the overall stress (eq \ref{eq:macro_stress}).
Theodorou and Suter\cite{Macromolecules_19_379} have observed atomic-level stresses of the same order of magnitude 
as those reported here, in the case of well-relaxed configurations of glassy atactic polypropylene.

Atomic-level stresses are sensitive to the topology of bonded systems.\cite{Macromolecules_19_379}
Aliphatic \ce{CH2} and aromatic \ce{CH} united atoms are under compression 
($p < 0$), while aliphatic CH and aromatic C united atoms are under tension ($p > 0$). 
The same observations can be made for the nanocomposite systems.
Fullerene carbon atoms, in the case of nanocomposite systems, are characterized by a broad $p$ distribution centered 
around zero. The moments of the distributions are reported in 
Table S3 of the Supporting Information to the present paper.
In all cases, the distributions are slightly skewed to the left, meaning the peak of the distribution lies 
on the right of its mean value. However, skewness values are close to zero, implying nearly Gaussian distributions of 
atomic level hydrostatic pressures.
The spread of the $p$ distribution reflects variations in density of the local environments. 
The $p$ distributions for aliphatic CH groups and aromatic carbons are considerably broader than those of aliphatic 
\ce{CH2} and aromatic \ce{CH} groups, implying a variety of environments experienced by the atoms belonging to the 
former groups.
The addition of fullerenes to PS does not seem to affect the atomic-level hydrostatic pressure distributions. 
Different species experience different shear stresses, as evidenced from Figure S3 and Table S4 of the Supporting 
Information to the present paper. 
Aromatic carbons are characterized by the broadest distribution of atomic-level von Mises shear stresses. 
This can be attributed to the exposure of the protruding phenyls to the surroundings of a chain, which creates a
highly asymmetric local environment. 
On the contrary, aliphatic \ce{CH2} interaction sites experience a considerably lower average shear stress, due to 
their shielded position in the backbone of the chain. 
The atomic shear-stress distributions are all skewed to the right (this is also the shape observed by Theodorou and 
Suter\cite{Macromolecules_19_379}) displaying extended tails to the right. Fullerene carbon atoms exhibit high atomic 
von-Mises shear stress, as expected, due to the stiff intramolecular potential which keeps them rigid.

\subsection{Distribution of local stresses}
In addition to the atomic-level stresses, we wish to obtain coarse-grained local stresses per Voronoi cell to examine the 
influence of the dispersed particles on the stress distribution in the material. There are several possibilities to 
compute the stress tensor in amorphous materials on intermediate scales. A widely accepted approach has been to 
partition the simulation cell into subvolumes, with only those atoms residing in each subvolume contributing to the 
local stress tensor of the region.\cite{PhysRevLett_93_175501}
The stress tensor within a region can be computed according to eq \ref{eq:macro_stress}, where the sum runs only over 
those atoms that are in the given region at a particular time:\cite{PhysRevE_81_011804} 
\begin{equation}
\boldsymbol \sigma_{\rm cell} = \frac{1}{V_{\rm cell}} \sum_{i \in {\rm cell}} V_i \boldsymbol \sigma_i
\label{eq:pcell_definition}
\end{equation}
This scheme converges to the macroscopic stress, eq \ref{eq:macro_stress}, when the simulation box is taken as a single
domain.
The ``local hydrostatic pressure'', $p_{\rm cell}$, and the ``local von Mises shear stress'', $\tau_{\rm cell}$, 
can be calculated at a local level using eq \ref{eq:atomic_p_definition} and eq \ref{eq:atomic_von_Mises}, 
respectively.

\begin{figure}
   \includegraphics[width=0.45\textwidth]{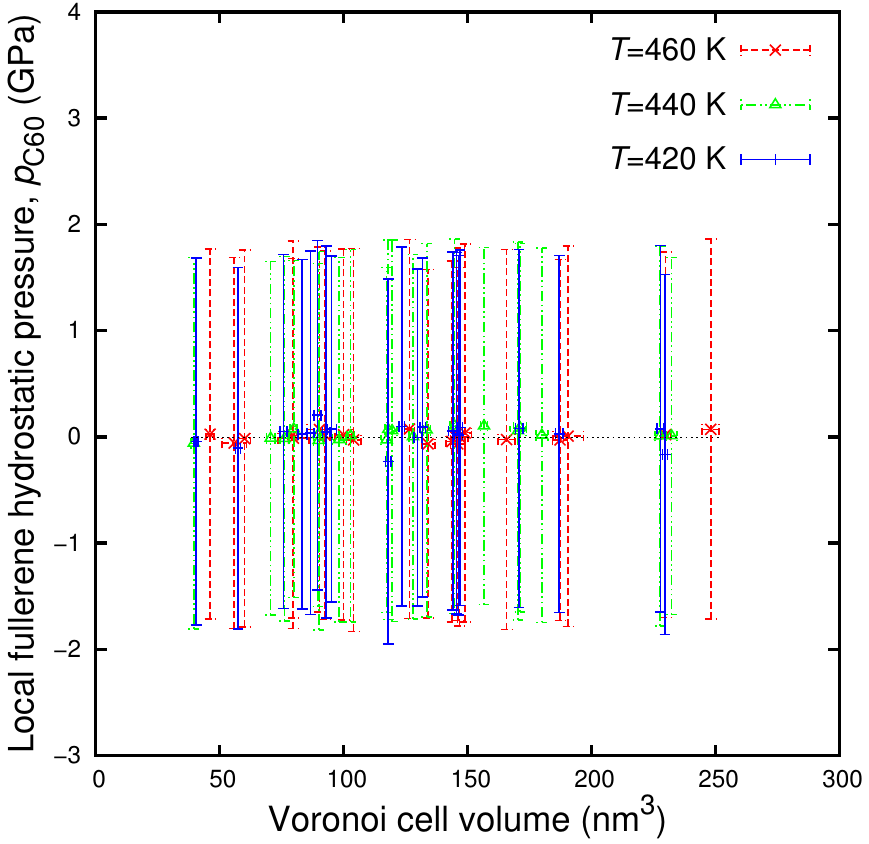}
   \caption{Local hydrostatic pressure of fullerenes, $p_{\rm C60}$, as a function of the volume of the Voronoi cell
   they belong to. Error bars represent the standard deviation of the distribution obtained by analyzing 
   1200 configurations at each temperature, obtained every 20 ps of MD simulation.}
   \label{fig:fullerene_hydrostatic}
\end{figure}

\begin{figure}
   \includegraphics[width=0.45\textwidth]{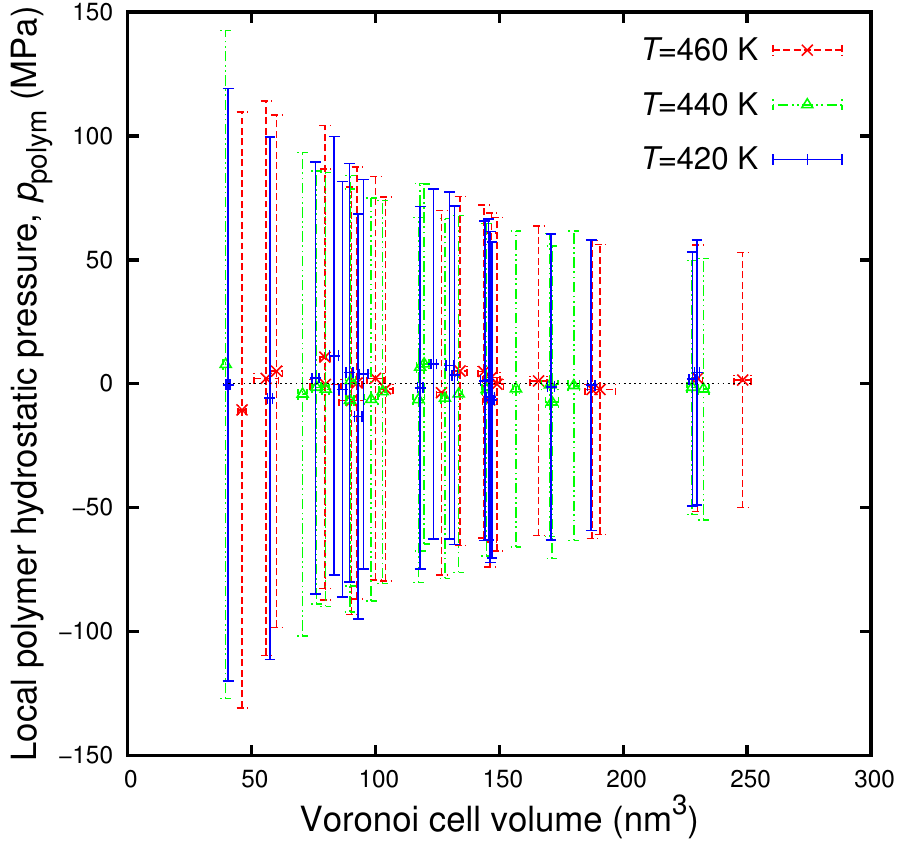}
   \caption{Local hydrostatic pressure of the polymeric atoms, $p_{\rm polym}$, as a function of the volume of the 
   Voronoi cell they belong to. Error bars represent the standard deviation of the distribution obtained by analyzing 
   1200 configurations at each temperature, obtained every 20 ps of MD simulation.}
   \label{fig:polymer_hydrostatic}
\end{figure}

The influence of confinement on the local hydrostatic pressure, $p_{\rm cell}$, as defined in eq 
\ref{eq:pcell_definition}, is examined in Figures \ref{fig:fullerene_hydrostatic} and \ref{fig:polymer_hydrostatic}. 
We first analyze the local pressure experienced by the fullerene atoms, $p_{\rm C60}$, summing the stresses only 
for the 60 atoms constituting each fullerene ( \ref{fig:fullerene_hydrostatic}). Then we study the local 
hydrostatic pressure experienced by the surrounding polymer atoms, $p_{\rm polym}$. These quantities, when added, 
yield the negative of the box pressure, 
$\left(\sum_{i \in cells}\left(p_{{\rm C60},i}V_{{\rm C60},i} + p_{{\rm polym},i}V_{{\rm polym},i}\right)\right)/3V = -p$.  
$V_{{\rm C60},i}$ is the sum of the volumes of the small (atomic) Voronoi polyhedra around the carbon atoms 
constituting a fullerene. By construction, this contains the hollow space in the fullerene, i.e. $V_{{\rm C60},i}$ is 
approximately equal to the volume of the fullerene sphere. On the other hand, $V_{{\rm polym},i}$ is the sum of the 
small atomic Voronoi volumes of polymer atoms contained in a larger Voronoi polyhedron around a fullerene obtained from 
the tessellation of the simulation box based on \ce{C60} centers. 
The volume of the large Voronoi cell, i.e. the one used for quantifying confinement, serves as the abscissa of both  
Figures \ref{fig:fullerene_hydrostatic} and \ref{fig:polymer_hydrostatic}.
The number of atoms per cell does not remain constant throughout the simulation but fluctuates slightly depending on 
the local particle density and the shape (number of faces) of the Voronoi cell.
Inspecting the local hydrostatic pressures, we see in Figures \ref{fig:fullerene_hydrostatic} and 
\ref{fig:polymer_hydrostatic} that the pressure distributions are almost symmetrically centered around the macroscopic
set point pressure.
The standard deviation of the distribution of local pressures becomes larger, as the volume of the Voronoi cell 
decreases. This fact is expected in the case of $p_{\rm polym}$ ( \ref{fig:polymer_hydrostatic}), 
where, as the length scale of the observation becomes
smaller, increasing fluctuations are observed.\cite{PhysRevE_81_011804}
On the contrary, in the case of $p_{\rm C60}$ ( \ref{fig:fullerene_hydrostatic}), the dispersion of local 
pressures is not affected by the size of the Voronoi cell the fullerenes occupy, 
since a constant number of atoms (60) contribute to the local hydrostatic pressure.
Fullerenes experience nearly zero hydrostatic pressure with a rather narrow distribution around its mean value. 
Figures \ref{fig:fullerene_hydrostatic} and \ref{fig:polymer_hydrostatic} indicate that the local hydrostatic pressures
are not sensitive to the temperature, in the melt state. 


\begin{figure}
   \includegraphics[width=0.45\textwidth]{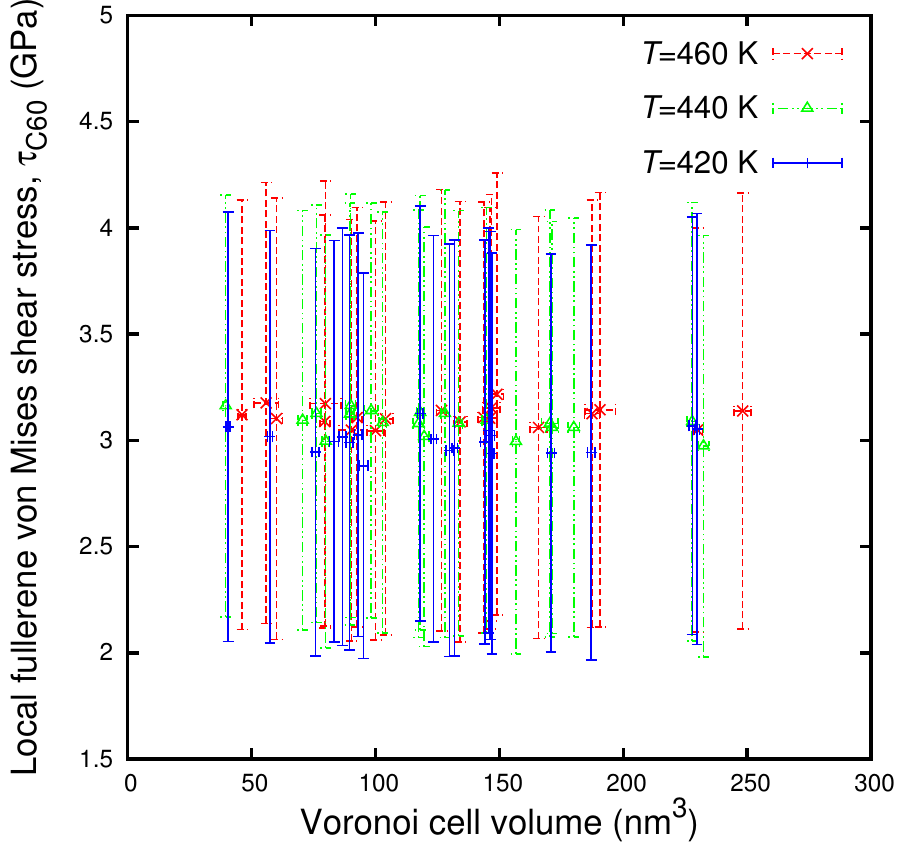}
   \caption{Local von Mises shear stress of the fullerenes, $\tau_{\rm C60}$, as a function of the volume of the 
   Voronoi cell they belong to. Error bars represent the standard deviation of the distribution obtained by analyzing 
   1200 configurations at each temperature, obtained every 20 ps of MD simulation.}
   \label{fig:fullerene_vonMises}
\end{figure}

\begin{figure}
   \includegraphics[width=0.45\textwidth]{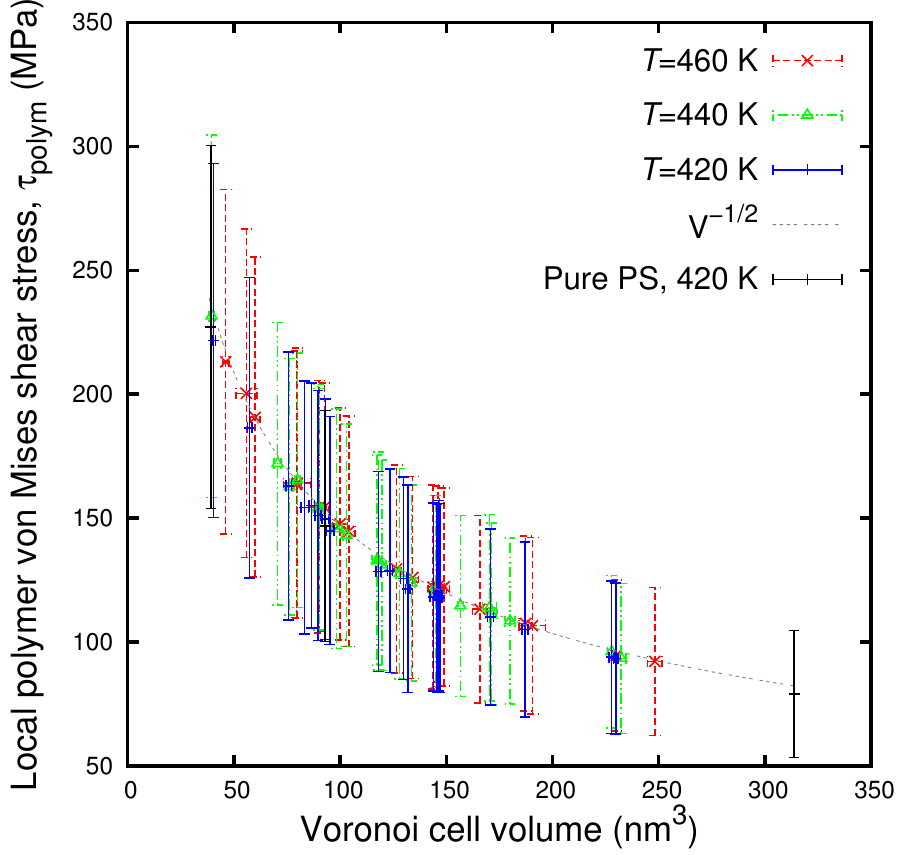}
   \caption{Local von Mises shear stress of the polymeric atoms, $\tau_{\rm polym}$, as a function of the volume of 
   the Voronoi cell they belong to. Error bars represent the standard deviation of the distribution obtained by analyzing 
   1200 configurations at each temperature, obtained every 20 ps of MD simulation. The dotted line shows the 
   $V^{-1/2}$ dependence.}
   \label{fig:polymer_vonMises}
\end{figure}

Figures \ref{fig:fullerene_vonMises} and \ref{fig:polymer_vonMises} present the local von Mises stresses of fullerenes 
and surrounding polymeric atoms, respectively.
As evident from  \ref{fig:fullerene_vonMises}, fullerenes experience strong shear stresses, which seem to
be independent of the volume the Voronoi cell which they occupy. The average fullerene von Mises shear stress is 
approximately 3 GPa, irrespectively of the simulation temperature.
This high value may be attributed to the stiff intramolecular potential which makes fullerenes behave as rigid bodies. 
On the contrary, the local von Mises shear stress of the polymer occupying a Voronoi cell is not fixed. It is a function
of the volume of the cell, as presented in  \ref{fig:polymer_vonMises}. 
Shear stresses become more pronounced, as the volume of the reference domain becomes smaller. As the length-scale of 
observation increases, fluctuations cancel each other yielding lower shear stress. At the length scale of the simulation
box, shear stresses almost vanish. This length-scale dependence of von Mises shear stress has been also found for pure
PS. In both cases (pure and composite), the von Mises shear stress scales as the inverse square root of the volume. 
This scaling can be envisioned as the dependence of an equilibrium fluctuation quantity on the observation length-scale.
The length-scale dependence of the shear stresses will be the focus of future work.
By a careful look at  \ref{fig:polymer_vonMises}, it can be observed that the addition of fullerenes shifts 
the von Mises shear stresses of polymeric atoms to slightly higher values, retaining the same length-scale dependence.

\section{Summary and Conclusions}
In this paper, we have outlined a strategy for simulating monodisperse long-chain atactic polystyrene nanocomposite 
melts at two interconnected levels of description: a coarse-grained one, wherein each diad along the chains is 
represented as a single interaction site, or ``superatom'', and a detailed one, employing a united-atom model 
representation for the polymer. Results from both levels, applied to a nanocomposite of high importance, have been 
presented. We focus on high molecular weight PS melts with fullerene (\ce{C60}) molecules dispersed at a mass fraction 
of 1\% , specifications identical to systems already studied experimentally.\cite{NanoLett_8_1061}

A state-of-the-art Monte Carlo builder has been developed which can build polymeric chains of arbitrary geometry in 
heavily constrained environments. It is based on the well-known quasi-Metropolis scheme of Theodorou and 
Suter\cite{Macromolecules_18_1467} for generating amorphous configurations in a bond-by-bond fashion. The original 
source code has been extended by an on-the-fly minimization strategy for driving the insertion of monomers in 
constrained environments. The minimizer generates a set of minima, from which one is selected following a quasi-Metropolis
procedure. 

The coarse-grained simulations of the PS-\ce{C60} composite were based on the PS model of Milano and 
M\"uller-Plathe.\cite{JPhysChem_B_109_18609} The coarse-grained model was equilibrated at $500\;{\rm K}$ using 
connectivity-altering Monte Carlo moves along with flips, rotations, reptations and concerted 
rotations.\cite{Macromolecules_40_3876} 
Chain conformations obtained through these coarse-grained simulations were found to be equilibrated at all length 
scales. Chain dimensions, as predicted from the mean square end-to-end distance ( \ref{fig:ete_vs_nu}) and 
from the mean square radius of gyrations ( \ref{fig:rg_vs_mw}) were found to be in excellent agreement with 
available experimental evidence. 
Chain conformations of the nanocomposite systems were found to be similar to the ones of the neat polymeric systems. 
No effect of the dispersion of \ce{C60}s has been observed at the coarse-grained level. 

Well-equilibrated melt configurations sampled by coarse-grained MC were reverse-mapped to the atomistic level. 
In continuation to the previous efforts on reverse mapping of coarse-grained polystyrene, for the first 
time a rigorous and systematic approach is presented, which yields atomistic configurations exhibiting characteristics
in excellent agreement with experimental measurements. Our approach encompasses three stages: (a) a quasi-Metropolis 
procedure for the re-introduction of the atomistic sites by selecting their locations from a set of candidate ones. (b)
A local Metropolis Monte Carlo simulation where flips, ring rotations and configurationally biased regrowths of 
atomistic sites are used.\cite{Macromolecules_40_3876} This is the key step in order to guide the reverse-mapping 
process into a conformationally reasonable subspace of the configuration space of the detailed model. (c) an energy 
minimization step using the full atomistic forcefield with gradual introduction of non-bonded interactions. During the 
whole reverse-mapping procedure, \ce{CH2} sites (centers of CG sites) are kept fixed, preserving the well-equilibrated 
coarse-grained configurations. The distributions of torsion angles in the reverse-mapped configurations (Figures 
\ref{fig:meso_torsion_distributions}(a) and \ref{fig:racemo_torsion_distributions}(a)) are indistinguishable from those 
of 80-mer configurations equilibrated directly via MD using the same united-atom model. Moreover, configurations 
obtained from reverse mapping exhibit populations of \textit{trans}, \textit{gauche} and \textit{gauche-bar} 
conformations in favorable agreement with NMR data and RIS calculations. Also, Ramachandran plots for successive torsion angles along
the chain backbone (Figures \ref{fig:meso_torsion_distributions}(b,c,d) and 
\ref{fig:racemo_torsion_distributions}(b,c,d)) exhibit symmetric \textit{tg} and \textit{gt} peaks and extremely low
percentages of \textit{t\=g} and \textit{\=g t} sequences. Torsion angle distributions constitute a stringent test for 
reverse-mapped structures, in which our reverse-mapping scheme fully succeeds. 

The ultimate goal of our work, i.e. the study of PS-\ce{C60} dynamics at the segmental and local levels, is accomplished
via analyzing long MD trajectories of our well-equilibrated reverse-mapped structures. Our simulation results generally
indicate that the addition of \ce{C60} to PS leads to slower segmental dynamics (as estimated by characteristic times 
extracted from the decay of orientational time-autocorrelation functions of suitably chosen vectors). The characteristic 
times found by fitting the $P_1(t)$ function of the orientation of the center of mass of the phenyl rings with respect
to the chain backbone suggest an increase of the bulk $T_{\rm g}$ of around $1\;{\rm K}$, upon the addition 
of \ce{C60}s at a concentration of 1\% by weight ( \ref{fig:p1_ring_com_wlf}). 
This observation is in favorable agreement with DS measurements of Kropka et al.\cite{NanoLett_8_1061} who
reported a $T_{\rm g}$ shift of $1\;{\rm K}$. The same conclusion can be reached by studying the $P_2(t)$ function of
the orientation of C-H bonds. Despite the fact that we employ a united-atom model for our MD simulations, the 
introduction of hydrogen atoms at a post-processing step yields reasonable dynamics for C-H bonds in good agreement 
with spin-lattice relaxation and solid echo NMR measurements on molten polystyrene ( \ref{fig:p2_aliphatic_wlf}). 
Again, nanocomposite systems are found to exhibit slightly slower dynamics than their neat polymer counterparts. 

We then employ a space discretization in order to study the effect of \ce{C60}s on segmental dynamics at a 
local level. Each fullerene
serves as the center of a Voronoi cell, whose volume is determined by the neighboring fullerenes. Backbone carbons lying 
in every cell are tracked throughout the MD trajectory and their mean-squared displacement is measured for the time they 
reside in the same cell. Overall mean-squared displacement of backbone atoms is found to be smaller in the presence of 
fullerenes, than in bulk PS ( \ref{fig:local_msd_480}). 
However, atoms moving in smaller (more confined) Voronoi cells exhibit faster motion 
than the atoms moving inside larger Voronoi cells. 
This can be correlated with the increased rotational diffusion of fullerenes, as the volume of the Voronoi cell becomes
smaller. These observations drive us to envision fullerenes as nanoscopic millstones, 
inducing shearing stresses on their environment and thus forcing the polymeric chains to move. The dynamic
heterogeneity caused by the addition of fullerenes exhibits strong temperature dependence, getting larger as the 
temperature is lowered.

Finally, we study the atomic-level and local stresses which are present in our systems. Each united atom is characterized 
by a distribution of atomic stresses whose shape and position reflect its chemical nature, connectivity, and geometric 
disposition within the system.
The addition of fullerenes causes an imperceptible broadening of the atomic-level stress distributions, 
implying a slightly higher stress heterogeneity for the composite material. This observation is further elaborated 
by the estimation of local stresses, at the level of Voronoi cells around individual fullerenes, into which the system 
has been partitioned. Dispersed fullerenes are characterized by large shear stresses, probably due to their 
intramolecular forcefield which reflects 
their rigidity. The local von Mises shear stress of the polymer has been found to scale as the inverse square root of 
the volume of the material used for the calculation, as expected from fluctuation theory. 
Local von Mises stresses are very similar between the neat polymer and the nanocomposite, being slightly higher in the 
latter. They are also rather insensitive to temperature.

Our analysis of polymer atom mean square displacement and fullerene rotational motion implies that fullerenes greatly 
amplify the dynamic heterogeneity of the molten atactic polystyrene, at the same time slowing down its overall dynamics 
slightly.
Further work is required to establish the effect of the addition of nanoparticles on terminal relaxation and viscosity.
In a future work we will focus on the effect of temperature and glass transition on the 
local stresses in the glassy state. 

\begin{acknowledgement} 
Part of this work was funded by the European Union through the project COMPNANOCOMP under grant 
number 295355. G.V. wants to thank the Alexander S. Onassis Public Benefit Foundation for a doctoral 
scholarship. 
We are grateful to Dr. Dora Spyriouni for sharing her PS MC code with us and thoroughly answering all our questions. 
Generous allocation of computer time on clusters of the School of Chemical Engineering of the National Technical 
University of Athens, expertly maintained by Prof. Andreas Boudouvis and his collaborators, is gratefully 
acknowledged. Computer time provided on clusters of the Department of Mechanical Engineering of the Technische 
Universiteit Eindhoven, maintained by Leo H.G. Wouters, is also acknowledged.
Code development was greatly facilitated by using hardware donated by nVIDIA to the authors through nVIDIA Academic 
Partnership Program. 
Last but not least, fruitful discussions with Prof. Polykarpos Pissis (NTUA), Dr. Loukas Peristeras (Scienomics),
Dr. Evangelos Voyiatzis (TU Darmstadt) and dr. ir. Lambert C.A. van Breemen (TU Eindhoven) are gratefully acknowledged. 
\end{acknowledgement} 

\begin{suppinfo}
Hydrogen reconstruction schematic, orientational autocorrelation functions fitting and atomic-level
stress distributions.
\end{suppinfo}

\bibliography{dynamics_manuscript}

\end{document}